\documentclass[journal]{IEEEtran}
\IEEEoverridecommandlockouts
\usepackage{amsmath,amsfonts}
\usepackage{array}
\usepackage{textcomp}
\usepackage{stfloats}
\usepackage{url}
\usepackage{verbatim}
\usepackage{graphicx}
\usepackage{cite}
\usepackage{microtype}
\usepackage{graphicx}
\usepackage{subcaption}
\usepackage{booktabs}
\usepackage{hyperref}
\usepackage{multicol}
\usepackage{multirow}
\usepackage{caption}
\usepackage{wrapfig}
\usepackage{xspace}
\usepackage{amsmath}
\usepackage{mathtools}
\usepackage{amsthm}
\usepackage{enumitem}
\usepackage{tikz}

\newenvironment{smitemize}{
  \begin{itemize}[topsep=1pt, partopsep=0pt, itemsep=1pt, parsep=0pt, leftmargin=10pt, itemindent=1pt]
}{\end{itemize}}

\usepackage{xcolor}

\usepackage{amsmath,amsfonts,bm}

\def\eqref#1{equation~\ref{#1}}

\def\1{\bm{1}}

\DeclareMathAlphabet{\mathsfit}{\encodingdefault}{\sfdefault}{m}{sl}
\SetMathAlphabet{\mathsfit}{bold}{\encodingdefault}{\sfdefault}{bx}{n}

\usepackage{algorithmic}
\usepackage[linesnumbered, ruled,vlined]{algorithm2e}

\newcommand{\xvbox}[2]{\makebox[#1][l]{#2}}
\SetAlFnt{\footnotesize}

\SetAlCapSty{xAlCapSty}

\SetCommentSty{xCommentSty}

\SetNlSty{mynlfont}{}{}

\usepackage{scalerel}
\usetikzlibrary{svg.path}
\definecolor{orcidlogocol}{HTML}{A6CE39}
\tikzset{
  orcidlogo/.pic={
    \fill[orcidlogocol] svg{M256,128c0,70.7-57.3,128-128,128C57.3,256,0,198.7,0,128C0,57.3,57.3,0,128,0C198.7,0,256,57.3,256,128z};
    \fill[white] svg{M86.3,186.2H70.9V79.1h15.4v48.4V186.2z}
                 svg{M108.9,79.1h41.6c39.6,0,57,28.3,57,53.6c0,27.5-21.5,53.6-56.8,53.6h-41.8V79.1z M124.3,172.4h24.5c34.9,0,42.9-26.5,42.9-39.7c0-21.5-13.7-39.7-43.7-39.7h-23.7V172.4z}
                 svg{M88.7,56.8c0,5.5-4.5,10.1-10.1,10.1c-5.6,0-10.1-4.6-10.1-10.1c0-5.6,4.5-10.1,10.1-10.1C84.2,46.7,88.7,51.3,88.7,56.8z};
  }
}

\newcommand\orcidicon[1]{\href{https://orcid.org/#1}{\mbox{\scalerel*{
\begin{tikzpicture}[yscale=-1,transform shape]
\pic{orcidlogo};
\end{tikzpicture}
}{|}}}}

\newcommand{\paragraphX}[1]{\vskip 4pt \noindent \textit{#1} \hskip .05in}

\usepackage[color=green]{todonotes}

\newcommand{\system}{{MADAR}\xspace}

\begin{document}

\title{\system: Efficient Continual Learning for Malware Analysis with Distribution-Aware Replay}


\author{
\IEEEauthorblockN{Mohammad Saidur Rahman~\orcidicon{0000-0001-6673-171X}}\IEEEauthorrefmark{4}\IEEEauthorrefmark{2}, 
\IEEEauthorblockA{University of Texas at El Paso, 
\texttt{msrahman3@utep.edu}}\\
\IEEEauthorblockN{Scott Coull~\orcidicon{0009-0003-6921-1842}}\IEEEauthorrefmark{4},
\IEEEauthorblockA{Google,
\texttt{scottcoull@google.com}}\\
\IEEEauthorblockN{Qi Yu~\orcidicon{0000-0002-0426-5407}}\IEEEauthorrefmark{2}, 
\IEEEauthorblockA{
Rochester Institute of Technology, 
\texttt{qyuvks@rit.edu}}\\
\and
\IEEEauthorblockN{Matthew Wright~\orcidicon{0000-0002-8489-6347}}\IEEEauthorrefmark{2}, 
\IEEEauthorblockA{
Rochester Institute of Technology,
\texttt{matthew.wright@rit.edu}}
}

\maketitle

\begin{abstract}

Millions of new pieces of malicious software (i.e., malware) are introduced each year. This poses significant challenges for antivirus vendors, who use machine learning to detect and analyze malware, and must keep up with changes in the distribution while retaining knowledge of older variants. Continual learning (CL) holds the potential to address this challenge by relaxing the requirements of the incremental storage and computational costs of regularly retraining over all the collected data. Prior work, however, shows that CL techniques, which are designed primarily for computer vision tasks, fare poorly when applied to malware classification. To address these issues, we begin with an exploratory analysis of a typical malware dataset, which reveals that malware families are heterogeneous and difficult to characterize, requiring a wide variety of samples to learn a robust representation. Based on these findings, we propose $\underline{M}$alware $\underline{A}$nalysis with $\underline{D}$istribution-$\underline{A}$ware $\underline{R}$eplay (MADAR), a CL framework that accounts for the unique properties and challenges of the malware data distribution. Through  extensive evaluation on large-scale Windows and Android malware datasets, we show that MADAR significantly outperforms prior work. This highlights the importance of understanding domain characteristics when designing CL techniques and demonstrates a path forward for the malware analysis domain.
\end{abstract}



\paragraphX{\textbf{Resources.}}
The code and datasets of this paper are available at: \textcolor{blue}{\url{https://github.com/IQSeC-Lab/MADAR}}.


\section{Introduction}

Advances in machine learning have significantly enhanced the detection and classification of malicious software, achieving notable success across various domains such as Windows executables~\cite{malwareguard, transcendingtranscend, chen2023continuous}, PDFs~\cite{maiorca2012pattern}, and Android applications~\cite{arp2014drebin, cade, abusnaina2022systematically}. Traditional models, trained on static datasets, are typically expected to perform well on new data under the assumption of a constant data distribution. However, in reality, both malicious (\emph{malware}) and benign (\emph{goodware}) software evolve continuously, necessitating regular model updates to adapt to changes in data distribution and maintain effectiveness. For example, the AV-TEST Institute reports approximately 450,000 new malware samples daily~\cite{av-test}, while VirusTotal processes over one million unique submissions each day~\cite{virustotal}. This scale creates immense challenges in training and even storing all the samples.

Training a malware classification model solely on new data can lead to \emph{catastrophic forgetting} (CF)~\cite{french1999catastrophic, robins1995catastrophic}, where previously learned information is forgotten, resulting in increased misclassification and allowing attackers to evade detection with older malware strains, known as {\em Retrograde Malware Attacks (RMA)} (see Section~\ref{sec:threat_model}). 
As such, anti-virus vendors must deal with difficult trade-offs: (i) removing older samples from the training set, risking exposure to revived older malware; (ii) ignoring newer samples, risking failure to detect emerging trends; (iii) reducing the frequency of retraining, compromising accuracy during intervals between updates; or (iv) incurring significant effort and cost to frequently retrain on the combined new and older samples. These challenges highlight the need for agile and adaptive malware classification techniques capable of learning incrementally and responding to the dynamic malware landscape.

Continual learning (CL) provides a promising solution to this problem by enabling models to adapt to new data without requiring the retention of large datasets or frequent retraining~\cite{malcl,wangunified, tamil, BIR, icarl, aljundi2019gradient}. By addressing catastrophic forgetting, CL techniques ensure that models remain effective and efficient in the face of evolving malware distributions. While designs for CL have been extensively studied in the context of computer vision~\cite{gr,hsu2018re,BIR}, there are very few such studies in the malware classification domain~\cite{continual-learning-malware, malcl}. Rahman et al.~\cite{continual-learning-malware} observed that CL techniques originally developed for computer vision problems fail to deliver acceptable performance in malware classification, due in part to the strong semantics of malware features and the high level of heterogeneity found in the  malware ecosystem.

In this study, we first delve into the complexities of malware data distributions using the EMBER dataset~\cite{ember}~\footnote{The recently released EMBER 2024 dataset~\cite{ember2024} is outside the scope of this study, as it was not available at the time the experiments were conducted.}. Our analysis highlights the heterogeneity in malware, both between and even within \emph{families}, or groups of related malware. Leveraging this insight, we devise \system\ -- $\underline{M}$alware $\underline{A}$nalysis with $\underline{D}$istribution-$\underline{A}$ware $\underline{R}$eplay, a replay-based continual learning strategy that accounts for heterogeneity and achieves improved malware classification performance. In particular, \system\ replays a mix of representative samples and novel samples (i.e., outliers) to enhance the model's ability to retain knowledge and identify new malware variants despite memory constraints. Our techniques employ Isolation Forests (IF)~\cite{if} to identify critical novel samples, either directly through the raw feature vectors (\system), or through the use of the hidden representations of the model for a more compact representation (MADAR$^\theta$). For both of these approaches, we consider two mechanisms to control how the number of replay samples is chosen, which we refer to as {\em Ratio} and {\em Uniform} strategies.

We evaluate these techniques with comprehensive experiments on two large-scale datasets across three CL scenarios representative of real-world malware analysis tasks, covering domain incremental learning (Domain-IL), class incremental learning (Class-IL), and task incremental learning (Task-IL). These datasets include the well-known EMBER dataset~\cite{ember}, containing one million examples of Windows executables, and a new benchmark dataset of Android malware from the AndroZoo repository~\cite{AndroZoo} specifically created to explore CL scenarios. Our experimental results across these datasets demonstrate that \system\ is effective and outperforms prior state-of-the-art continual learning methods when confronted with realistic distribution shifts in malware data.


In summary, the contributions of this study are:
\begin{itemize}
    \item We provide an exploratory analysis of the heterogeneity of malware distributions and show how it creates unique challenges for continuous learning.

    \item We develop a large-scale, realistic Android malware benchmark dataset covering all three CL scenarios -- Domain-IL, Class-IL, and Task-IL.
    
    \item In Domain-IL scenarios, we show that \system\ performs much better than prior CL techniques. On the AndroZoo dataset, for example, \system\ comes within 0.4\% average accuracy of the  retraining baseline using just 50K training samples versus 680K for full retraining.

    \item \system\ is also effective in Class-IL scenarios, where it consistently outperforms all prior methods over a wide range of budgets. With a budget of 20K training samples on EMBER, \system\ gets an average accuracy of $85.8\%$ versus $66.8\%$ for the best method from prior work.
    
    \item For Task-IL, \system\ outperforms all prior methods across all memory budget configurations for both the EMBER and AndroZoo datasets. For example, in the AndroZoo dataset, the \system-U variant of MADAR achieves an average accuracy of 98.7\% (within 0.1\% of full retrain) with a budget of only 20K replay samples (versus approximately 250K for full retrain).
    
\end{itemize}

Through these contributions, this study stands as a significant advancement in continual learning for malware classification, highlighting the importance of understanding the domain distribution to effectively combat catastrophic forgetting.

\section{Threat Model}
\label{sec:threat_model}

\begin{figure}[!t]
    \centering
    \includegraphics[width=0.80\linewidth]{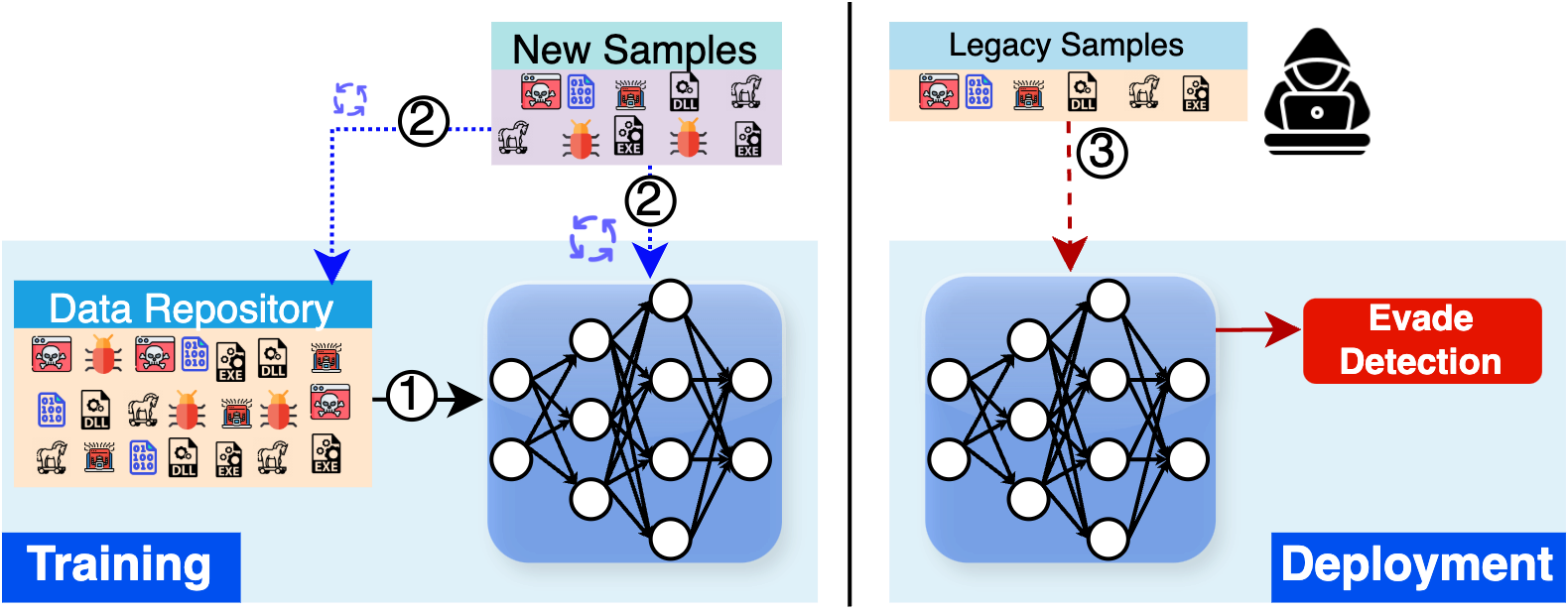}
    \vspace{-0.2cm}
    \caption{Retrograde Malware Attack (RMA).}
    \label{fig:CFthreatModel}
    \vspace{-0.4cm}
\end{figure}

\begin{figure}[!t]
    \centering
    \includegraphics[width=0.90\linewidth]{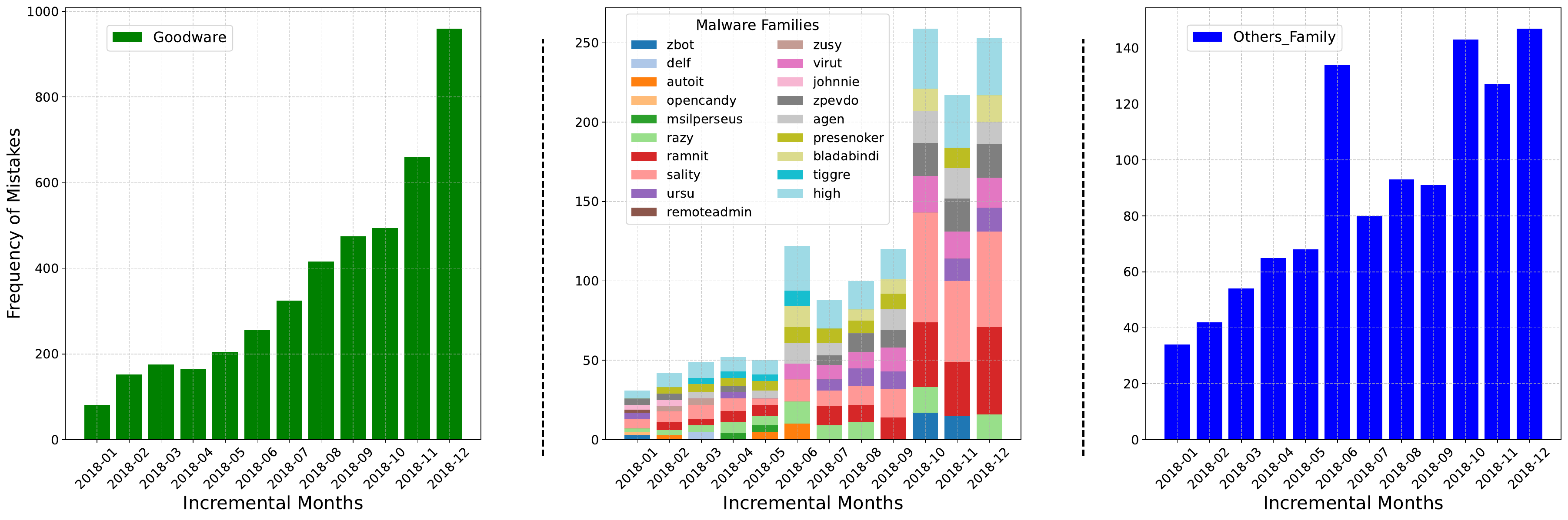}
    \vspace{-0.2cm}
    \caption{Frequencies of forgotten goodware, top malware families, and malware with unlabeled families (i.e., \texttt{others\_family}) across incremental monthly learning episodes for EMBER dataset.}
    \label{fig:forgotten_frequencies}
    \vspace{-0.4cm}
\end{figure}

The \textit{Retrograde Malware Attack (RMA)} describes an attack scenario where adversaries can exploit catastrophic forgetting in machine learning-based malware detection systems. RMA occurs when a system, updated incrementally with only {\em new samples}, loses the ability to recognize older malware, allowing attackers to reintroduce legacy or slightly modified variants that can evade detection.

As shown in Figure~\ref{fig:CFthreatModel}, the attack can be realized in the deployment phase of the detection system:

\begin{itemize}
    \item \textbf{Initial Training and Deployment ({\Large \textcircled{\normalsize 1}}):} 
    The system is trained on an initial dataset of malware and benign software and deployed for classification.

    \item \textbf{Incremental Updates and Forgetting ({\Large \textcircled{\normalsize 2}}):} 
    As new samples emerge, the model is retrained with only the latest data. This process leads to catastrophic forgetting, where older malware signatures are no longer retained, reducing detection accuracy for previously known threats. In addition, false positives increase as the model struggles to differentiate between benign and malicious samples, leading to the misclassification of legitimate software. Figure~\ref{fig:forgotten_frequencies} illustrates the frequencies of forgotten goodware, top malware families, and malware with unlabeled families (i.e., \texttt{others\_family}) over incremental monthly learning episodes for EMBER dataset~\cite{ember}, demonstrating the extent of catastrophic forgetting in the model.

    \item \textbf{Retrograde Malware Attack (RMA) ({\Large \textcircled{\normalsize 3}}):} 
    Attackers exploit this limitation by \textit{reintroducing forgotten malware or slightly modified versions}, bypassing detection as the model has lost prior knowledge. The success of this attack increases with each incremental update if no mechanisms are in place to retain past information.
\end{itemize}

This attack presents a challenge for CL-based malware detection as maintaining detection accuracy for both new and previously known threats is critical. We address this by proposing \system~a replay-based CL framework that mitigates catastrophic forgetting and improves robustness against RMA.

\section{Related Work}
\label{related_work}

\subsection{Replay in Continual Learning}
The fundamental challenge in developing a CL system is addressing catastrophic forgetting, and one of the widely studied methods to overcome forgetting is {\em replay}. In general, replay techniques complement the training data for each new task with older data that are representative of the tasks observed by the model so far. These techniques can further be classified into one of three subcategories -- exact replay, generative replay, and compressed replay. 


\paragraphX{Exact Replay.} 
Selecting and utilizing replay samples in exact replay involves determining a memory budget, denoted as $\mathcal{B}$. Finding the optimal way to choose $\mathcal{B}$ remains an open research question~\cite{aljundi2019gradient,chaudhry2019tiny}. Exact-replay techniques are designed to choose replay samples from previously learned data to be combined with new samples for retraining. The goal of these techniques is to maximize the performance with minimal replay samples~\cite{er,agem,icarl, smith2024adaptive}. 



\paragraphX{Generative/Pseudo Replay.} 
Generative or pseudo-replay strategies are designed to replicate the original data~\cite{lwf,gr,BIR, malcl}. These techniques either generate a representative of the original data using a separate generative model or generate pseudo-data by using an earlier model's predictions as soft labels for training subsequent models.

\subsection{CL in Malware and Related Domains.} 
Despite extensive work in CL, very few studies have ventured into applying CL in the realm of malware. To the best of our knowledge, Rahman et al.\cite{continual-learning-malware} were the first to explore CL for malware classification. They concluded that existing CL methods fall short in tackling forgetting in malware classification systems due to differences in the underlying nature of the data distribution shifts that occur in practice versus those explored in the computer vision domain. Malware representations leverage tabular features with strong semantic constraints that limit the space of feasible samples, and within that space, samples exhibit a high level of heterogeneity. Replay-based techniques are found to perform better compared to other approaches in this setting. Another recent work, MalCL~\cite{malcl}, introduces a generative replay-based continual learning approach for malware classification using a generative adversarial network (GAN). It achieves state-of-the-art performance compared to prior generative replay methods. However, its evaluation is restricted to the Class-IL scenario.

Furthermore, Chen et al. proposed to combine contrastive learning with active learning to continuously train Android malware classifiers~\cite{chen2023continuous}. They focus on the detection of {\em concept drift}, rather than overcoming forgetting.In addition, some studies have used \emph{online learning} for malware classification, which deals with adding new samples as they are observed but does not directly address catastrophic forgetting~\cite{droidevolver}. 
Another CL domain in cybersecurity is network intrusion detection (NID), looking for malicious activity based on network packets. Amalapuram et al. explored a replay-based CL technique that incorporates class-balancing reservoir sampling and perturbation assistance for parameter approximation NID~\cite{channappayya2024augmented}. Another recent work explored semi-supervised CL for NID in a class incremental setting~\cite{amalapuram2024spider}. We note that NID is a different domain with different data characteristics than malware. Further, these works do not focus on reducing CF.



\section{Preliminaries}
\label{preliminaries}

\subsection{CL Scenarios for Malware Classification}

\begin{figure}
    \centering
    \includegraphics[width=\linewidth]{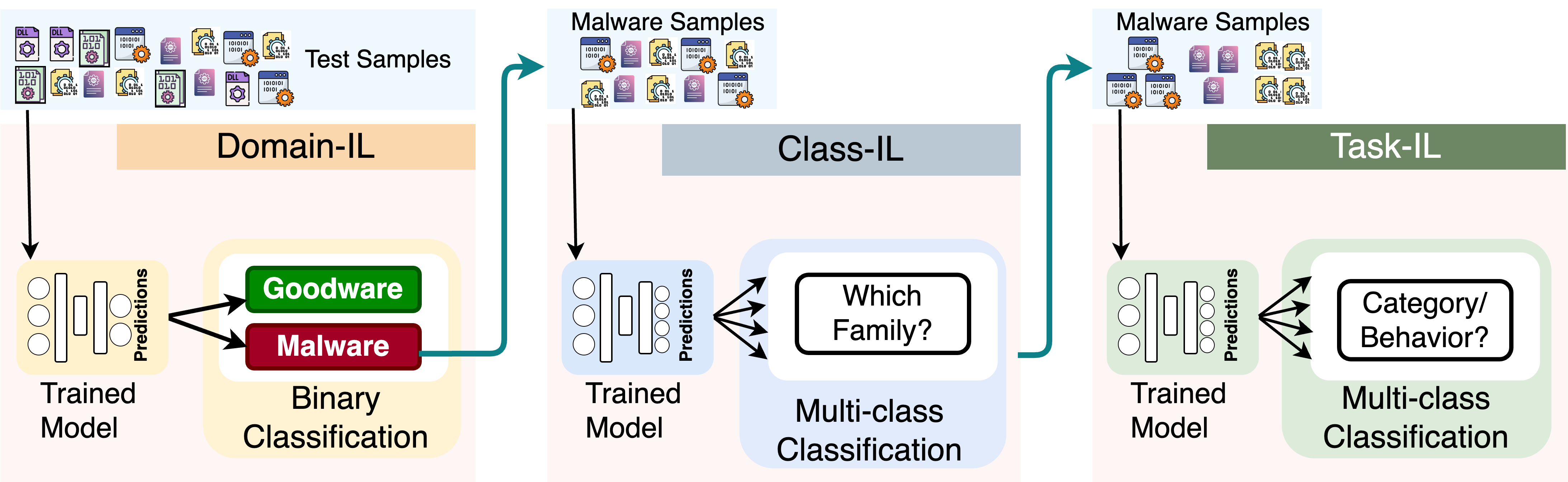}
    \caption{CL scenarios in a typical malware analysis pipeline.}
    \label{fig:cl-scenarios}
    \vskip -0.4cm
\end{figure}


Continual Learning (CL) is categorized into three scenarios: Domain Incremental Learning (Domain-IL), Class Incremental Learning (Class-IL), and Task Incremental Learning (Task-IL)~\cite{van2022three}. In this work, we demonstrate how the  three CL scenarios naturally fit into a typical malware analysis pipeline (see Figure~\ref{fig:cl-scenarios}). The first step in the pipeline is to determine whether a test sample is malware or goodware. Next, the pipeline identifies the specific family of the detected malware, which is formulated as a multi-class classification problem. Finally, the pipeline classifies the broader category or behavior of the malware, such as adware or ransomware, which is also a multi-class classification problem. 
 


\subsubsection{Domain-IL}
The primary challenge in malware classification lies in distinguishing between goodware and malware. Each day, VirusTotal receives one million never-before-seen samples~\cite{virustotal}, highlighting the persistent and evolving nature of software, known as \emph{concept drift}~\cite{chen2023continuous}. This evolution underscores the importance of rapidly integrating these new samples into operational systems to maintain effective protections against new threats. In addition, with the continual emergence of new benign software programs and the massive class imbalance in practice (i.e., significantly more goodware than malware), it is of utmost importance to not increase the false positive rate of the classifiers.

In this adversarial context, attackers might deploy older malware to evade detection by systems that have {\em forgotten} their signatures, necessitating a balance between adapting to new threats and preventing catastrophic forgetting. To address this, we segment our Domain-IL datasets into monthly tasks for EMBER and yearly tasks for AZ to mirror natural temporal shifts in the threat distribution.


\subsubsection{Class-IL} 


Another significant task in malware analysis involves classifying malware into families, which are groups of programs with substantial code overlap and similar functionalities, as recognized by experts~\cite{zhu2020measuring}. For instance, the \texttt{zeus} banking trojan has evolved into 556 variants across 35 families, including \texttt{citadel} and \texttt{gameover}. Labeling new samples often relies on consensus from multiple anti-virus engines and occurs when a significant set of similar samples forms a new family~\cite{kantchelian2015better,zhu2020measuring}. In our incremental multi-class model, we start with known malware families and add new ones as they emerge, continuously adjusting and assessing the model across all known classes.


\subsubsection{Task-IL}
In malware analysis, leveraging insights from additional methods can prove beneficial. This may involve identifying the broader malware category (e.g., adware, ransomware, etc.), malware behaviors~\cite{maliciousbehavior}, or the infection vector (e.g., phishing, downloader, etc.). Task-IL encapsulates this concept of constrained tasks, where the introduction of a new task may symbolize a new category or behavior. This event occurs less frequently than adding a new family, as seen in Class-IL, yet it poses a genuine issue. Unlike Class-IL, the task identity is provided to the model at test time, significantly simplifying the problem. In malware, this could mean learning the task identity from a separate model, manual analysis, or field reports of the malware's behavior. However, as our datasets do not possess naturally defined tasks, we partition our dataset into tasks comprising an equal number of independent and non-overlapping classes to act as a proxy to new behaviors, following common practice in the CL literature~\cite{van2022three,BIR}. In other words, a given task would be to perform family classification among one subset of families, and the subset that each sample belongs to is known to the classifier. The model is expected to be able to handle multiple tasks at once, and new tasks are being added during each experiment.

\subsection{Dataset}\label{sec:dataset}

In this study, we use large-scale Windows and Android malware datasets: EMBER~\cite{ember}, a Windows malware dataset from prior work, recognized as a benchmark for malware classification, and two new Android malware datasets derived from AndroZoo~\cite{AndroZoo}, specifically assembled for this research.

\subsubsection{Windows PE Files}

For our experiments, we leverage the EMBER 2018 dataset, containing features from one million Windows Portable Executable (PE) files, predominantly scanned in 2018~\footnote{\url{https://github.com/elastic/ember}}. The dataset comprises 400K goodware and 350K malware, with the rest labeled as unknown. EMBER provides a diverse array of 2,381 hand-crafted features, covering general file information, header data, import/export functions, and section details. Notably, these features capture strong semantic concepts that have a limited space of feasible settings, outside of which the executable does not actually run.

In our Class-IL experiments, we focused on 2018 malware samples from 2,900 families. After filtering out families with fewer than 400 samples, we narrowed the remaining samples down to the top 100 families, leaving 337,035 samples for analysis. For Domain-IL, we included both goodware and malware from the entire year of 2018 for binary classification, excluding unknown samples.

\subsubsection{Android APK Files}

Additionally, we collected two datasets from AndroZoo~\cite{AndroZoo} (AZ) for our experiments: AZ-Domain for Domain-IL and AZ-Class for Class-IL and Task-IL. These datasets contain Android APK files, and both use a 9:1 ratio of goodware to malware to reflect real-world class imbalance. Following the practice of prior work~\cite{droidevolver}, the malware samples are selected with a VirusTotal detection count of $>= 4$. The AZ-Domain dataset includes 80,690 malware and 677,756 goodware samples from 2008 to 2016. We divided the AZ-Domain dataset into non-overlapping yearly training and testing sets. The AZ-Class dataset consists of 285,582 samples from 100 Android malware families, each with at least 200 samples.

We extracted Drebin features~\cite{arp2014drebin} from the apps for both datasets. These features cover various aspects of app behavior, including hardware access, permissions, app component names, filtered intents, restricted API calls, used permissions, suspicious API calls, and network addresses. Again, we note that these capture strong semantic concepts from the operation of the application. The training sets of AZ-Domain and AZ-Class have 3,858,791 and 1,067,550 features, respectively. We processed the test datasets to match the training feature sets and reduced dimensionality by filtering features with low variance ($<0.001$) using \texttt{scikit-learn}’s \texttt{VarianceThreshold}. This resulted in final feature dimensions of 1,789 for AZ-Domain and 2,439 for AZ-Class, respectively.


\subsection{Model Selection and Implementation}
\label{sec:modelSelectionTraining}

We use a multi-layer perceptron (MLP) model for malware classification, similar to the model used by Rahman et al.~\cite{continual-learning-malware}, for experiments with the EMBER dataset. For the AZ dataset, we developed a new MLP model with five fully-connected layers, quite similar to the MLP used for EMBER. This model uses the Adam optimizer with a learning rate of $0.001$, and batch normalization and dropout for regularization. 


The implementation of the output layer varies among Domain-IL, Class-IL, and Task-IL scenarios. Domain-IL operates as a series of binary classification tasks over 12 months for EMBER, and over 9 years for the AZ dataset, with two output units in each case: malicious and benign. In Class-IL, the output layer comprises units -- one for each malware family. Output units are active only if they correspond to family that have been seen by that point in the experiment. Class-IL begins with an initial set of 50 families in the first task and progressively adds five more families in each of the remaining 10 tasks for both EMBER and AZ datasets. In Task-IL, only the output units of the families in the current task are active. Both the EMBER and AZ-Class datasets divide the families equally into 20 tasks, with each task containing five families. 


\subsection{Baselines and Metric}
\label{sec:baselines}

We adopt two baselines for comparison: {\em None} and {\em Joint}.  {\em None} sequentially trains the model on each new task without any CL techniques, serving as an informal minimum baseline. By contrast, {\em Joint} uses all new and prior data for training at each step, acting as an informal maximum baseline. Despite its resource demands, {\em Joint} ensures strong performance throughout the dataset. 
We also introduce an additional baseline -- Global Reservoir Sampling -- which provides an unbiased sampling of the underlying class distributions and serves as a strong point of comparison for our distribution-aware approach.

\paragraph*{Global Reservoir Sampling (GRS)}
\label{sec:grs}

GRS simply selects samples at random from a global stored data pool~\cite{vitter1985random,zhang2017deeper}. Given a memory budget $\beta$, GRS randomly picks $\beta$ samples from a data pool $\mathcal{P}$, with each incremental learning task contributing to the pool. 
If $\beta \geq \mathcal{P}$, GRS selects all the available samples in $\mathcal{P}$. 
Rahman et al.~\cite{continual-learning-malware} investigated GRS -- which they refer to as Partial Joint Replay -- only for Domain-IL scenario of EMBER dataset. In this work, we present a deeper investigation of GRS in both Domain-IL and Class-IL scenarios with both EMBER and AZ datasets. 



\paragraph*{Global average accuracy ($\overline{AP} \in [0, 100]\%$)}
To maintain consistency with prior work, we present results using \emph{global average accuracy} as the primary metric for our evaluations~\cite{continual-learning-malware, BIR, icarl}. Note that we conducted a subset of evaluations using other metrics, such as F1 score, precision, and recall, which are not included in this paper. The conclusions remain unchanged for all of these metrics.

Let $P_{i,j}$ be the accuracy of the model on the test set
of task $T_j$, $j \leq i$, after 
continually training the model on tasks $1~\text{to}~i$. Then the average accuracy $AP$ at task $T_i$ is defined as $AP_i = \frac{1}{i} \sum_{j=1}^i P_{i,j}$. For $N$ total tasks, the global average accuracy $\overline{AP}$ over all tasks is computed as $\overline{AP} = \frac{1}{N} \sum_{i=1}^{N} AP_i * 100$.



\subsection{Training and Evaluation Protocol}


A continual learning (CL) model is sequentially trained to learn tasks from $t_1, t_2, ..., t_T$, each with its distinct data distribution $p(x,y|t_i)$. The goal is to adapt to new tasks without forgetting the old ones. CL training involves three sets of parameters: shared parameters ($\theta_{s}$) across all tasks, old task-specific parameters ($\theta_{0}$), and new task parameters ($\theta_{n}$)~\cite{lwf}. The {\em Joint} training benchmark trains the model with all the available training samples up to the current task and optimizes all these parameters simultaneously; however, it incurs incremental storage and training costs. In contrast, CL training strives to optimize and update $\theta_{s}$ and $\theta_{n}$, while maintaining $\theta_{0}$ in a relatively fixed state for each new task $t_{n}$. However, updating any of the shared weights $\theta_{s}$ risks confusing the classifier when faced with older data, as those classification decisions depend not only on $\theta_{0}$ but also on $\theta_{s}$. CL training typically boasts significantly faster speeds and far less storage requirements than {\em Joint} training, thus permitting more frequent model retraining to adapt to evolving data distributions or other requirements.


In our evaluations, we use a non-overlapping hold-out set corresponding to each task. For example, the AZ-Domain dataset contains 8 years of training samples from 2008 to 2016, resulting in 9 hold-out sets, one for each year. A CL model is evaluated on all the hold-out sets up to the current task; formally, 
the model is evaluated on tasks $t_i$ to $t_{T}$, for $1 \leq i \leq T$, after it been trained on the current task $t_{T}$. In this work, each set of experiments is performed around 10-15 times with different random parameter initializations. We use PyTorch on a CentOS-7 machine with an Intel Xeon processor, 40 CPU cores, 128GB RAM, and four GeForce RTX 2080Ti GPUs, each with 12GB memory.
\section{Exploratory Analysis of EMBER}
\label{sec:exploratoryAnalysis}

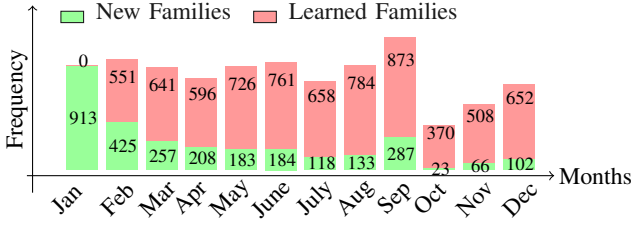
\begin{figure}[!t]
\tiny
\begin{minipage}[c]{\linewidth}
\centering
\begin{tikzpicture}[scale=1.5]

    \tikzstyle{every node}=[font=\small]
    \def\scalefactor{0.001}
    
    \def\NewFamilies{{913, 425, 257, 208, 183, 184, 118, 133, 287, 23, 66, 102}}
    \def\LearnedFamilies{{0, 551, 641, 596, 726, 761, 658, 784, 873, 370, 508, 652}}

    \def\Months{{"Jan","Feb","Mar","Apr","May","Jun","Jul","Aug","Sep","Oct","Nov","Dec"}}
    
    \foreach \x [count=\xi] in {0,...,11}{
        \pgfmathsetmacro\NewFamily{\NewFamilies[\x]}
        \pgfmathsetmacro\LearnedFamily{\LearnedFamilies[\x]}
        
        \pgfmathsetmacro\NewHeight{\NewFamily*\scalefactor}
        \pgfmathsetmacro\LearnedHeight{(\NewFamily + \LearnedFamily)*\scalefactor}
        
        \pgfmathsetmacro\position{\xi*0.35}
        
        \fill[green!40] (\position,0) rectangle (\position+0.275,\NewHeight);
        \fill[red!40] (\position,\NewHeight) rectangle (\position+0.275,\LearnedHeight);
        
        \node at (\position+0.15,\NewHeight/2) {\scriptsize\NewFamily};
        \ifnum\LearnedFamily = 0
            \node at (\position+0.15,\LearnedHeight*1.05) {\scriptsize\LearnedFamily};
        \fi
         \ifnum\LearnedFamily > 0
            \node at (\position+0.15,\LearnedHeight/1.2) {\scriptsize\LearnedFamily};
        \fi
    }

    \node[rotate=45] at (0.35,-0.25) {Jan};
    \node[rotate=45] at (0.8,-0.25) {Feb};
    \node[rotate=45] at (1.15,-0.25) {Mar};
    \node[rotate=45] at (1.45,-0.25) {Apr};
    \node[rotate=45] at (1.80,-0.25) {May};
    \node[rotate=45] at (2.15,-0.25) {June};
    \node[rotate=45] at (2.55,-0.25) {July};
    \node[rotate=45] at (2.90,-0.25) {Aug};
    \node[rotate=45] at (3.25,-0.25) {Sep};
    \node[rotate=45] at (3.55,-0.25) {Oct};
    \node[rotate=45] at (3.95,-0.25) {Nov};
    \node[rotate=45] at (4.35,-0.25) {Dec};

    \draw[->] (0.0,-0.05) -- (4.65,-0.05) node[right] {Months};
    \draw[->] (0.05,-0.2) -- (0.05,1.45) node[rotate=90, midway, above] {Frequency};

    \draw [black!90, fill=green!40] (.3,1.3) rectangle (0.5,1.4) node[right, xshift=0.1cm] {\small New Families};
    \draw [black!90, fill=red!40] (2.0,1.3) rectangle (2.20,1.4) node[right, xshift=0.1cm] {\small Learned Families};
    
\end{tikzpicture}
\vskip -0.1cm
\caption{New and already learned families in each task.}
\label{fig:ember_frequency_families}
\end{minipage}%
\vspace{-0.3cm}
\end{figure}

In this section, we provide an analysis of the EMBER dataset, which sheds light on the distribution across various families and tasks, aiding in selecting representative samples for replay. We identified 2,899 unique malware families within a subset of EMBER, and an additional 11,433 samples lacking clear family labels were assigned the label {\em Other}.

We investigate the prevalence of malware families over time, distinguishing between recurring and newly identified families each month in Figure~\ref{fig:ember_frequency_families}. Unlike many datasets used in CL research, we see significant churn in the representation of families over time. Of the 913 families seen in January, for example, only 551 are seen in February, while 425 new families emerge. This churn indicates a potential issue in training data continuity, which may aggravate catastrophic forgetting and underscores the need for different CL strategies in the malware domain. Table~\ref{ember_task_family_stat} shows the number of goodware and malware samples in each month, as well as the number of families. Generally, each family has its own distinct distribution pattern, and together these patterns make up the total distribution of malware for a particular month. 

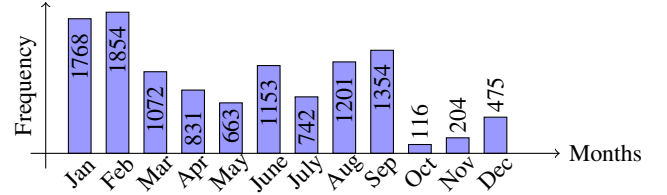
\begin{figure}[t]
\begin{minipage}[c]{\linewidth}
\centering
\begin{tikzpicture}[scale=1.]
    \tikzstyle{every node}=[font=\small]
    
    \newcommand{\scalefactor}{0.001}
    
    \draw[fill=blue!40] (0,0) rectangle (0.3,1768*\scalefactor); 
    \draw[fill=blue!40] (0.5,0) rectangle (0.8,1854*\scalefactor); 
    \draw[fill=blue!40] (1,0) rectangle (1.3,1072*\scalefactor); 
    \draw[fill=blue!40] (1.5,0) rectangle (1.8,831*\scalefactor); 
    \draw[fill=blue!40] (2,0) rectangle (2.3,663*\scalefactor); 
    \draw[fill=blue!40] (2.5,0) rectangle (2.8,1153*\scalefactor); 
    \draw[fill=blue!40] (3,0) rectangle (3.3,742*\scalefactor); 
    \draw[fill=blue!40] (3.5,0) rectangle (3.8,1201*\scalefactor); 
    \draw[fill=blue!40] (4,0) rectangle (4.3,1354*\scalefactor); 
    \draw[fill=blue!40] (4.5,0) rectangle (4.8,116*\scalefactor); 
    \draw[fill=blue!40] (5,0) rectangle (5.3,204*\scalefactor); 
    \draw[fill=blue!40] (5.5,0) rectangle (5.8,475*\scalefactor); 
    
    \node[rotate=45] at (0.15,-0.25) {Jan};
    \node[rotate=45] at (0.65,-0.25) {Feb};
    \node[rotate=45] at (1.15,-0.25) {Mar};
    \node[rotate=45] at (1.65,-0.25) {Apr};
    \node[rotate=45] at (2.15,-0.25) {May};
    \node[rotate=45] at (2.65,-0.25) {June};
    \node[rotate=45] at (3.15,-0.25) {July};
    \node[rotate=45] at (3.65,-0.25) {Aug};
    \node[rotate=45] at (4.15,-0.25) {Sep};
    \node[rotate=45] at (4.65,-0.25) {Oct};
    \node[rotate=45] at (5.15,-0.25) {Nov};
    \node[rotate=45] at (5.65,-0.25) {Dec};
    
    \node[rotate=90] at (0.15,1768*\scalefactor-0.5) {1768};
    \node[rotate=90] at (0.65,1854*\scalefactor-0.5) {1854};
    \node[rotate=90] at (1.15,1072*\scalefactor-0.5) {1072};
    \node[rotate=90] at (1.65,831*\scalefactor-0.5) {831};
    \node[rotate=90] at (2.15,663*\scalefactor-0.3) {663};
    \node[rotate=90] at (2.65,1153*\scalefactor-0.5) {1153};
    \node[rotate=90] at (3.15,742*\scalefactor-0.4) {742};
    \node[rotate=90] at (3.65,1201*\scalefactor-0.5) {1201};
    \node[rotate=90] at (4.15,1354*\scalefactor-0.5) {1354};
    \node[rotate=90] at (4.65,116*\scalefactor+0.3) {116};
    \node[rotate=90] at (5.15,204*\scalefactor+0.3) {204};
    \node[rotate=90] at (5.65,475*\scalefactor+0.3) {475};
    
    \draw[->] (-0.5,0) -- (6.5,0) node[right] {Months};
    
    \draw[->] (-0.3,-0.2) -- (-0.3,2) node[rotate=90, midway, above] {Frequency};
    
\end{tikzpicture}
\vskip -0.25cm
\caption{EMBER Malware samples without AV-Class labels.}
\label{fig:ember_noavclass}
\end{minipage}%
\vskip -0.4cm
\end{figure}

Worse, many malware samples do not have family labels at all (see Figure~\ref{fig:ember_noavclass}). Correctly labeling samples is a challenging endeavor and often takes time and expert knowledge~\cite{kantchelian2015better}, so this lack of labels matches real-world conditions. Family labels for malware are based on the {\em av-class} labels provided by the av-test engine~\cite{av-test}. Analysis of the \emph{Other}-labeled samples do not align with the labeled families, which shows that many of them indeed come from unknown families.

\begin{table}
\scriptsize
\centering
\caption{Number of samples and families per task in EMBER}
\begin{tabular}{l|c|c|c} 
\textbf{Task} & \textbf{\#of Goodware} & \textbf{\#of Malware} & \textbf{\#of Families}\\ 
\hline

January     &   29423   &   32491    &  913       \\
February    &   22915   &   31222    &  976      \\
March       &   21373   &   20152    &  898      \\
April       &   25190   &   26892    &  804      \\
May         &   23719   &   22193    &  909      \\
June        &   23285   &   25116    &  945      \\
July        &   24799   &   26622    &  776      \\
August      &   23634   &   21791    &  917      \\
September   &   26707   &   37062    &  1160      \\
October     &   29955   &   56459    &  393      \\
November    &   50000   &   50000    &  574      \\
December    &   50000   &   50000    &  754      \\

\bottomrule

\end{tabular}
\label{ember_task_family_stat}
\vskip -0.4cm   
\end{table}

Furthermore, our analysis shows that the prominent malware families change with the evolution of tasks. We identified the 10 most common malware families based on the frequency of samples from each family and found that the top 10 families vary across tasks. The most prevalent families at the beginning of the experiment (i.e., January) do not remain prevalent in later tasks (i.e., from February on). For example, the \texttt{emotet} malware family was the most consistent, appearing in 11 out of 12 tasks. The next most consistent families were \texttt{fareit} and \texttt{zusy}, appearing in eight and seven tasks, respectively. This indicates considerable concept drift in malware data, highlighting the need to regularly update classifiers.

\begin{figure}[!t]
\centering
\vskip -0.3cm
\includegraphics[scale=0.20,trim={0.5cm 0.8cm 0.5cm 0.5cm},clip]{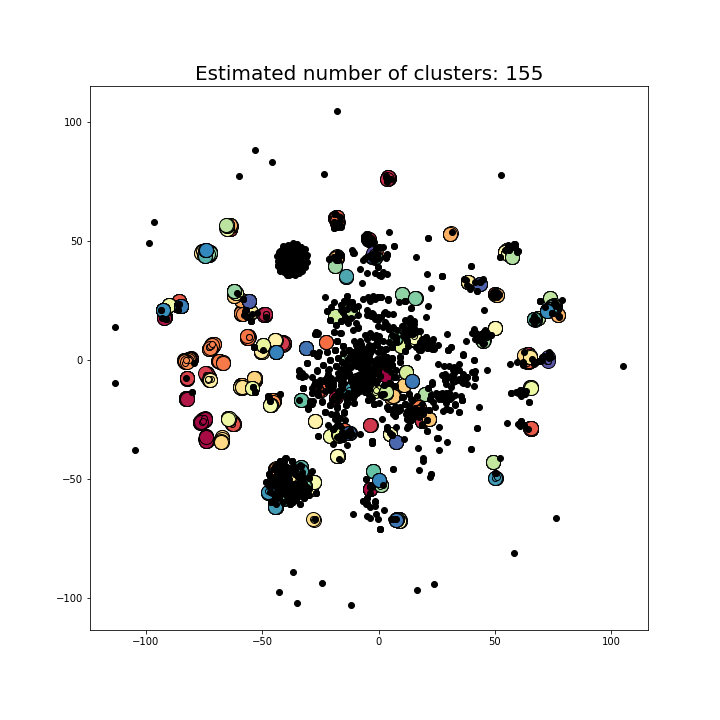}
\vskip -0.3cm
\caption{t-SNE projection of EMBER malware from Jan. 2018.}
\label{fig:ember_jan_mal}
\vskip -0.4cm
\end{figure}

In addition, many malware families display complex distributional patterns in feature space, making for additional heterogeneity within classes. Figure~\ref{fig:ember_jan_mal}, for example, shows a t-SNE projection of the EMBER features for all malware samples from January 2018. Each class (represented by color) is not clustered into a single well-defined region. Rather, the larger classes are split up into subsets that spread out in feature space. To accurately represent the malware distribution, it is thus important to select samples not only from each class, but also from multiple areas within each class.

Despite some degree of overlap among different families, it is possible to discern distinct clusters indicative of the heterogeneity across and even within malware families. It is vital to capture samples from each of these smaller clusters, including those belonging to minority families, to accurately represent the malware landscape for a specific task month. 
To further complicate the situation, it is often not possible to provide definitive family labels for a sample due to the inherent subjectivity involved in malware analysis, which results in a large, diverse set of additional unlabeled samples which must be considered~\cite{kantchelian2015better}. These attributes of the malware landscape make it difficult to characterize classes as contiguous regions of the feature space with relatively simple class boundaries. Rather, each class might only be understood as a collection of smaller pockets of the feature space that might be closer to pockets from other classes than the same class. This may explain why prior CL techniques designed for computer vision datasets are less effective when applied to the malware domain~\cite{continual-learning-malware}.




In light of these analysis, we propose that selecting samples based on families and variations within those families could more effectively capture the heterogeneity within the replay data distribution, potentially mitigating catastrophic forgetting.

\section{Distribution-Aware Replay}
\label{divreplay}

Here, we introduce the our proposed continual learning (CL) framework for malware classification with two distribution-aware replay variants: MADAR and MADAR$^\theta$. MADAR operates on the basis of the raw-feature space and MADAR$^\theta$ operates on the basis of the model's weights-space of raw-features.

\subsection{MADAR}
\label{subsec:ifs}



Building on the exploratory analysis in Section~\ref{sec:exploratoryAnalysis}, we postulate that stratified sampling, where replay samples are chosen based on their representation in malware families, may better preserve the model's stability versus random sampling as used in GRS. Moreover, we also seek to capture the heterogeneity \emph{within} each family's data distribution. 
We do this by selecting a balance between \emph{representative samples} that are close to each other and \emph{anomalous samples} that are separated. 
While any single anomalous sample is not as important to learn and remember as a single representative sample, a collection of anomalous samples helps to track the heterogeneity within a class.


{\em Isolation Forest (IF)}~\cite{if} is a technique for identifying outliers in high-dimensional data. IF uses decision trees to isolate anomalous data points based on the intuition that outliers are distinct and easy to separate from the rest of the data. An important parameter in IF is the contamination rate $C_{r}$, which represents the expected fraction of outliers in the data. We found that $C_{r} = 0.1$ works best and used it in all our experiments. The algorithm for \system\ in the Domain-IL setting is provided in Algorithm~\ref{alg:IFS}. The algorithm for \system\ in the Class-IL and Task-IL settings is provided in Algorithm~\ref{alg:IFS_Class_Task_IL}.

\begin{algorithm}[!t]
\small
\SetKwFunction{cumprod}{cumprod}
\SetKwFunction{length}{length}
\SetKwFunction{zeros}{zeros}
\SetKwFunction{ceil}{ceil}
\SetKwInOut{Input}{Input}
\SetKwInOut{Output}{Output}
\caption{~\system~in Domain-IL. 
\label{alg:IFS}}
\Input{
    $c$ -- Current Task number, $X_{c}, Y_{c}$ -- Samples and labels of $c$, 
    $\mathcal{P}$ -- Data pool, 
    $\beta$ -- Memory budget,  $\gamma$ -- Split of $\beta$ for malware and goodware, 
    $\Omega$ -- Split of anomalous/similar samples, 
    $\xi$ -- Ratio budgeting, $\Psi$ -- Uniform budgeting
}
  \BlankLine 

    \xvbox{14mm}{\texttt{\bf init~~} $\mathcal{P}$;}
    \xvbox{1mm}{\texttt{\bf init~~} $\mathcal{D} \leftarrow \{M_{f}: M_{c}\}$;}
    
    \If{$c = 0$}{
        \xvbox{20mm}{$\mathcal{P} \leftarrow X_{c}, Y_{c}$;}
        
        \xvbox{20mm}{$X_{good}, X_{mal} \leftarrow  \mathcal{P}$;} 
        
        \xvbox{20mm}{$\nabla \mathcal{D} \leftarrow X_{mal}$}
        
        \xvbox{20mm}{$X_{train},~Y_{train} \leftarrow X_{c}, Y_{c}$}
        }
        
    
    \Else{
        \xvbox{40mm}{$X_{good}, X_{mal} \leftarrow  \mathcal{P}$;} 
        
        \xvbox{22mm}{$\beta_M, \beta_G \leftarrow \beta~.~\gamma$;} 

        \xvbox{26mm}{$\beta_A$, $\beta_S \leftarrow \beta_M~.~\Omega$;} 
        \If{$\Psi$}{ 
            \xvbox{20mm}{$\mathcal{NF} \leftarrow |\mathcal{D}|$;} 
            \xvbox{37mm}{$\mathcal{B_{F}} \leftarrow \beta_M / \mathcal{NF}$;} 
        }

        \xvbox{28mm}{$R_{mal} \leftarrow [~]$;} 
        \BlankLine
        \For{$X_{f} \subseteq X_{mal}$ 
        }{
                \xvbox{20mm}{$\mathcal{F_{MC}} \leftarrow X_{f}$;}
                
                \If{$\xi$}{
                    \xvbox{20mm}{$\mathcal{MC} \leftarrow |\mathcal{D}|$;}
                    \xvbox{37mm}{$\mathcal{B_{F}} \leftarrow (\mathcal{F_{MC}} / \mathcal{MC}) * \beta_M$;}
                }
                
                \If{$\mathcal{F_{MC}} <= \mathcal{B_{F}} $}{
                    \xvbox{20mm}{$R_{mal}.\texttt{append}(X_{f})$;}
                }

                \Else {
                    $({A}_f, {S}_f) \leftarrow \texttt{IF}(X_{f}, \beta_A, \beta_S);$ 
                    {
                    }
                    
                    $R_{mal}.\texttt{append}({A}_f, {S}_f)$
                }
        }
        
        \xvbox{80mm}{$R_{good}  \leftarrow \texttt{sample}(X_{good}, \texttt{len}(R_{mal}))$;} 

        \xvbox{40mm}{$X_{replay}  \leftarrow (R_{good}, R_{mal})$;}
        
        \xvbox{40mm}{$Y_{replay}  \leftarrow ([0] * \texttt{len}(R_{good}), [1]*\texttt{len}(R_{mal}))$;}

        
        \xvbox{40mm}{$X_{train}  \leftarrow \texttt{concat}((X_{c}, X_{replay}))$;}

        \xvbox{20mm}{$Y_{train}  \leftarrow (\texttt{concat}((Y_{c}, Y_{replay}))$;}

        $\mathcal{P}.\texttt{append}(X_{c}, Y_{c})$; 
    $\nabla \mathcal{D} \leftarrow X_{mal}$;

    }
    \texttt{\bf return} $(X_{train},~Y_{train})$
\end{algorithm}




\subsubsection*{\bf Procedure}
We now describe \system using the framework of Domain-IL; the process is similar for Class-IL and Task-IL. 
The procedure begins by initializing a global data pool $\mathcal{P}$, containing both goodware and malware samples, and a dictionary $\mathcal{D}$ that tracks malware families and their frequencies in the data up to the current task.

For each new task $c$, \system divides the data into goodware ($X_{good}$) and malware ($X_{mal}$) subsets from $\mathcal{P}$, allocating memory budgets $\beta_M$ for malware and $\beta_G$ for goodware from the total memory budget $\beta$, based on a split ratio $\gamma$:
\begin{equation}
    \beta_G = \gamma \cdot \beta, \quad  \beta_M = (1 - \gamma) \cdot \beta. 
\end{equation}
For balanced datasets like EMBER, $\gamma = 0.5$ ensures an equal split between malware and goodware. For an imbalanced dataset, it is better to tune $\gamma$. Our Android malware (AZ) datasets, for example, have a 9:1 ratio of goodware to malware, so we use $\gamma=0.9$.

Before training for a new task, \system incrementally trains the classifier using a combination of new samples from the current task and replay samples from previous tasks. The replay samples include both goodware ($R_{good} \subset X_{good}$) and malware ($R_{mal} \subset X_{mal}$), with $R_{mal}$ sampled from specific malware families rather than randomly from all of $X_{mal}$.

For each family $f$, we set its \emph{family budget} $\mathcal{B}_f$---
the number of samples to select from $f$---using two sub-sampling variants: \emph{Ratio budgeting} and \emph{Uniform budgeting}.

\begin{smitemize}
    \item \textbf{Ratio Budgeting:} Select the number of samples from a family $f$ proportional to that family's representation in the dataset. The family budget $\mathcal{B}_f$ is
    $\mathcal{B}_f = \frac{|X_{f}|}{|X_{mal}|} \cdot \beta_M$,
    %
    where $|X_{f}|$ is the number of samples in family $f$, and $|X_{mal}|$ is the total number of malware samples. This strategy may be more suitable in binary classification, as it provides proportional representation of the malware families for training on the malicious class.
    
    \item \textbf{Uniform Budgeting:} In this method, the memory budget $\beta_M$ is uniformly distributed across all malware families:
        $\mathcal{B}_f = \frac{\beta_M}{|\mathcal{F}|}$,
    where $|\mathcal{F}|$ is the total number of malware families. 
    Compared with Ratio budgeting, Uniform budgeting may work well for multi-class classification to determine which family a sample belongs to, as it ensures better class balance.
\end{smitemize}

Within each family $f$, we further split the family budget $\mathcal{B}_f$ into two parts: representative samples $S_f$ and anomalous samples $A_f$, using IF, controlled by a split parameter $\alpha$:
\begin{equation}
    |S_f| = \alpha \cdot \mathcal{B}_f, \quad |A_f| = (1 - \alpha) \cdot \mathcal{B}_f 
\end{equation}
%
We found empirically that a balanced split ($\alpha = 0.5$) between representative and anomalous samples provides optimal performance. In this setup, the model learns equally the core class characteristics from representative samples 
and less common variations from anomalous samples. 

The malware replay set $R_{mal}$ is then constructed by combining the representative and anomalous samples from all malware families: 
\begin{equation}
    R_{mal} = \bigcup_{f \in F} \{ S_f \cup A_f \}. 
\end{equation}
The total replay set consists of both goodware and malware replay samples, which are then concatenated with the new task samples
to form the training set for the current task $c$.
After training, the data pool $\mathcal{P}$ is updated with the new task samples, \(\mathcal{P} \leftarrow \mathcal{P} \cup \{X_c, Y_c\}\), and the malware family dictionary $\mathcal{D}$ is updated to reflect the new frequencies of malware families in $X_{mal}$: \(\mathcal{D} \leftarrow \mathcal{D} + freq(X_{mal})\).

\begin{algorithm}[!t]
\small
\caption{\system~in Class-IL and Task-IL \label{alg:IFS_Class_Task_IL}}
\SetKwInOut{Input}{Input}
\SetKwInOut{Output}{Output}

\Input{
    $c$ -- Task number, $X_{c}, Y_{c}$ -- Malware samples and their family labels, 
    $\mathcal{P}$ -- Malware data pool, 
    $\beta$ -- Memory budget, 
    $\Omega$ -- Split of anomalous/similar samples, 
    $\xi$ -- Ratio budgeting, $\Psi$ -- Uniform budgeting
}

\xvbox{14mm}{\texttt{\bf init~~} $\mathcal{P}$;}
\xvbox{1mm}{\texttt{\bf init~~} $\mathcal{D} \leftarrow \{M_{f}: M_{c}\}$;}

\If{$c = 0$}{
    $\mathcal{P} \leftarrow X_{c}, Y_{c}$; 
    $X_{mal} \leftarrow \mathcal{P}$; 
    $\nabla \mathcal{D} \leftarrow X_{mal}$; 
    $X_{train}, Y_{train} \leftarrow X_{c}, Y_{c}$;
}

\Else{
    $X_{mal} \leftarrow \mathcal{P}$; 
    $\beta_A, \beta_S \leftarrow \beta \cdot \Omega$; 
    
    \If{$\Psi$}{
        $\mathcal{NF} \leftarrow \mathcal{D}$; 
        $\mathcal{B_F} \leftarrow \beta / \mathcal{NF}$;
    }
    
    $R_{mal} \leftarrow [~]$; 
    \For{$X_{f} \subseteq X_{mal}$}{
        $\mathcal{F_{MC}} \leftarrow X_{f}$; 
        \If{$\xi$}{
            $\mathcal{MC} \leftarrow \mathcal{D}$; 
            $\mathcal{B_F} \leftarrow (\mathcal{F_{MC}} / \mathcal{MC}) \cdot \beta$;
        }
        \If{$\mathcal{F_{MC}} \leq \mathcal{B_F}$}{
            $R_{mal}.\texttt{append}(X_{f})$;
        } \Else{
            $({A}_f, {S}_f) \leftarrow \texttt{IF}(X_{f}, \beta_A, \beta_S)$; 
            $R_{mal}.\texttt{append}({A}_f, {S}_f)$;
        }
    }
    
    $X_{replay} \leftarrow R_{mal}$; 
    $Y_{replay} \leftarrow ([1] \times \texttt{len}(R_{mal}))$; 

    $X_{train} \leftarrow \texttt{concat}(X_{c}, X_{replay})$; 
    $Y_{train} \leftarrow \texttt{concat}(Y_{c}, Y_{replay})$;
    
    $\mathcal{P}.\texttt{append}(X_{c}, Y_{c})$; 
    $\nabla \mathcal{D} \leftarrow X_{mal}$;
}
\textbf{return} $(X_{train}, Y_{train})$
\end{algorithm}

\subsection{MADAR$^\theta$}
\label{subsec:aws}

While Isolation Forest (IF) outperforms other distance-based anomaly detection techniques on high-dimensional data, it struggles with correlated features~\cite{puggini2018enhanced}. Let \( \mathbf{X} \in \mathbb{R}^D \) represent the input data, where \( D \) is the feature dimension. For example, the EMBER dataset has \( D = 2381 \), while the AZ datasets have \( D = 1789 \) for AZ-Domain and \( D = 2439 \) for AZ-Class, respectively. Instead of applying dimensionality reduction techniques such as PCA, we propose leveraging a neural network \( \mathcal{M} \) to generate compact feature representations \( \mathbf{Z} \in \mathbb{R}^d \), where \( d \ll D \). This approach is particularly advantageous in continual learning contexts, as it complements ongoing model development and adapts to evolving tasks~\cite{hayes2020remind,ostapenko2022foundational}.

To address these, we introduce MADAR$^\theta$, a variant of MADAR that leverages learned representations from a trained model \( \mathcal{M} \) to identify both representative and diverse samples effectively. 
For an input sample \( \mathbf{x} \), the model \( \mathcal{M} \) computes activations \( \mathcal{W}(\mathbf{x}) \), representing its internal state. Specifically, for a set of inputs \( \mathbf{X}_f \) belonging to a family \( f \), MADAR$^\theta$ extracts activations:
\begin{equation}
\Theta^\mathcal{L}(\mathbf{X}_f) = \{ \mathcal{W}^\mathcal{L}(\mathbf{x}) : \mathbf{x} \in \mathbf{X}_f \}, \quad \text{where} \quad \Theta^\mathcal{L}(\mathbf{X}_f) \in \mathbb{R}^d,
\end{equation}
from a chosen layer \( \mathcal{L} \) of the model.

These activations are analyzed using the Isolation Forest (IF) algorithm to identify anomalous activations \( \mathcal{A}_w \subseteq \Theta^\mathcal{L}(\mathbf{X}_f) \). The corresponding anomalous samples in the original input space are denoted as:
\begin{equation}
A_f = \{ \mathbf{x} \in \mathbf{X}_f : \mathcal{W}^\mathcal{L}(\mathbf{x}) \in \mathcal{A}_w \}.
\end{equation}
Non-anomalous samples are similarly sampled to form the set \( S_f \), ensuring a balanced and representative replay set:
\begin{equation}
S_f = \{ \mathbf{x} \in \mathbf{X}_f : \mathcal{W}^\mathcal{L}(\mathbf{x}) \notin \mathcal{A}_w \}.
\end{equation}

The replay samples \( A_f \cup S_f \) are then used in subsequent training episodes to maintain knowledge retention and mitigate catastrophic forgetting.

\paragraphX{Selection of the Layer \( \mathcal{L} \).}
The choice of \( \mathcal{L} \) is critical in MADAR$^\theta$ to ensure that feature representations are captured without interference from the model's final classification stage. Empirical testing revealed that the penultimate layers, denoted as \texttt{act4} for the EMBER dataset and \texttt{act5} for the AZ datasets, provide optimal results such that $\mathcal{L}_{\text{EMBER}} = \texttt{act4}, \quad \mathcal{L}_{\text{AZ}} = \texttt{act5}.$


\paragraphX{Adaptation of Batch Normalization.}
During the forward pass, batch normalization is omitted in MADAR$^\theta$ to preserve the heterogeneity of the activation distributions. While batch normalization typically stabilizes learning and improves generalization, it can homogenize activations, reducing the heterogeneity essential for identifying anomalies. This adaptation enhances the performance of MADAR$^\theta$, as confirmed by empirical results.


\paragraphX{Efficiency of MADAR$^\theta$.}
MADAR$^\theta$ is computationally efficient compared to MADAR applied directly to the input space. The hidden layers \texttt{act4} and \texttt{act5} have \( d = 128 \), significantly reducing the dimensionality from \( D \). As a result, MADAR$^\theta$ offers reduced computational complexity.

\section{Evaluation}


\begin{table*}[!t]
\small
\centering
\caption{Summary of Results for EMBER Domain-IL Experiments.}
\vspace{-0.2cm}
\label{tab:ember_DIL}

\begin{tabular}{p{1.1cm}|l|c|c|c|c|c|c|c} 


\multirow{2}{*}{\textbf{Group}} & \multirow{2}{*}{\textbf{Method}} & \multicolumn{7}{c}{\textbf{Budget}} \\ \cline{3-9}

&  & 1K & 10K & 50K & 100K & 200K & 300K & 400K \\ \midrule

\multirow{3}{*}{Baselines} 
& Joint  & \multicolumn{7}{c}{96.4$\pm$0.3} \\ 
& None   & \multicolumn{7}{c}{93.1$\pm$0.1} \\ 
& GRS    & 93.6$\pm$0.3 & 94.1$\pm$1.3 & 95.3$\pm$0.2 & 95.3$\pm$0.7 & 95.9$\pm$0.1 & 95.8$\pm$0.6 & 96.0$\pm$0.3 \\ 
\midrule

\multirow{4}{*}{\parbox{0.7cm}{Prior \\ Work}} 
& ER~\cite{er}     & 80.6$\pm$0.1 & 73.5$\pm$0.5 & 70.5$\pm$0.3 & 69.9$\pm$0.1 & 70.0$\pm$0.1 & 70.0$\pm$0.1 & 70.0$\pm$0.1 \\ 
& AGEM~\cite{agem}   & 80.5$\pm$0.1 & 73.6$\pm$0.2 & 70.4$\pm$0.3 & 70.0$\pm$0.1 & 70.0$\pm$0.2 & 70.0$\pm$0.1 & 70.0$\pm$0.1 \\ 
& GR~\cite{gr}     & \multicolumn{7}{c}{93.1$\pm$0.2} \\ 
& RtF~\cite{rtf}    & \multicolumn{7}{c}{93.2$\pm$0.2} \\ 
& BI-R~\cite{BIR}   & \multicolumn{7}{c}{93.4$\pm$0.1} \\ 
\midrule

\multirow{4}{*}{\system}      
& \system-R         & \textbf{93.7$\pm$0.1} & \textbf{94.7$\pm$0.1} & \textbf{95.4$\pm$0.1} & \textbf{95.3$\pm$0.6} & \textbf{96.0$\pm$0.1} & \textbf{96.1$\pm$0.1} & \textbf{96.1$\pm$0.1} \\ 
& \system-U         & \textbf{93.6$\pm$0.2} & 94.0$\pm$0.2 & 95.1$\pm$0.1 & \textbf{95.3$\pm$0.1} & 95.5$\pm$0.1 & 95.7$\pm$0.1 & 95.8$\pm$0.1 \\  \cline{2-9}
& MADAR$^{\theta}$-R & \textbf{93.6$\pm$0.1} & \textbf{94.4$\pm$0.3} & \textbf{95.3$\pm$0.2} & \textbf{95.8$\pm$0.1} & \textbf{96.1$\pm$0.1} & \textbf{96.1$\pm$0.1} & \textbf{96.1$\pm$0.1} \\ 
& MADAR$^{\theta}$-U & 93.5$\pm$0.2 & 94.1$\pm$0.2 & 94.9$\pm$0.1 & 95.2$\pm$0.2 & 95.6$\pm$0.1 & 95.7$\pm$0.1 & 95.7$\pm$0.1 \\ 

\bottomrule

\end{tabular}
\vspace{-0.2cm}
\end{table*}

\begin{table*}[!t]
\small
\centering
\caption{Summary of Results for AZ Domain-IL Experiments.}
\vspace{-0.3cm}
\label{tab:az_DIL}
\begin{tabular}{p{1.1cm}|l|c|c|c|c|c|c|c} 


\multirow{2}{*}{\textbf{Group}} & \multirow{2}{*}{\textbf{Method}} & \multicolumn{7}{c}{\textbf{Budget}} \\ \cline{3-9}

&  & 1K & 10K & 50K & 100K & 200K & 300K & 400K \\ \midrule

\multirow{3}{*}{Baselines} 
& Joint  & \multicolumn{7}{c}{97.3$\pm$0.1} \\ 
& None   & \multicolumn{7}{c}{94.4$\pm$0.1} \\ 
& GRS    & 95.3$\pm$0.1 & 96.4$\pm$0.1 & 96.9$\pm$0.1 & 97.1$\pm$0.1 & 97.1$\pm$0.1 & 97.2$\pm$0.1 & 97.2$\pm$0.1 \\ 
\midrule

\multirow{4}{*}{\parbox{0.7cm}{Prior \\ Work}} 
& ER~\cite{er}     & 40.4$\pm$0.1 & 40.1$\pm$0.1 & 41.1$\pm$0.2 & 42.6$\pm$0.1 & 44.0$\pm$0.1 & 45.9$\pm$0.1 & 48.6$\pm$1.1 \\ 
& AGEM~\cite{agem}   & 45.4$\pm$0.1 & 47.4$\pm$0.2 & 49.2$\pm$0.2 & 53.7$\pm$0.6 & 54.2$\pm$0.3 & 54.8$\pm$0.4 & 56.7$\pm$0.3 \\ 
& GR~\cite{gr}     & \multicolumn{7}{c}{93.3$\pm$0.4} \\ 
& RtF~\cite{rtf}     & \multicolumn{7}{c}{93.4$\pm$0.2} \\ 
& BI-R~\cite{BIR}     & \multicolumn{7}{c}{93.5$\pm$0.1} \\ 
\midrule

\multirow{4}{*}{\system}      
& \system-R         & \textbf{95.8$\pm$0.1} & \textbf{96.6$\pm$0.1} & \textbf{96.9$\pm$0.1} & \textbf{97.0$\pm$0.1} & \textbf{97.0$\pm$0.1} & \textbf{97.0$\pm$0.1} & \textbf{97.0$\pm$0.1} \\ 
& \system-U         & \textbf{95.7$\pm$0.1} & 95.5$\pm$0.1 & 95.2$\pm$0.2 & 95.2$\pm$0.1 & 95.4$\pm$0.1 & 95.8$\pm$0.2 & 96.3$\pm$0.2 \\ \cline{2-9}
& MADAR$^{\theta}$-R & \textbf{95.8$\pm$0.2} & \textbf{96.6$\pm$0.1} & \textbf{96.9$\pm$0.1} & \textbf{96.9$\pm$0.1} & \textbf{97.1$\pm$0.1} & \textbf{97.1$\pm$0.1} & \textbf{97.2$\pm$0.1} \\ 
& MADAR$^{\theta}$-U & 95.6$\pm$0.1 & 96.1$\pm$0.1 & 96.6$\pm$0.1 & 96.8$\pm$0.1 & \textbf{97.0$\pm$0.1} & \textbf{97.1$\pm$0.1} & \textbf{97.1$\pm$0.1} \\ 

\bottomrule

\end{tabular}
\vspace{-0.3cm}
\end{table*}

We present the results of our \system\ framework and MADAR$^\theta$ in the Domain-IL, Class-IL, and Task-IL scenarios using the EMBER and AZ datasets discussed in Section~\ref{sec:dataset}. To denote our techniques, we use the following abbreviations: {\bf \system-R} for \system-Ratio, {\bf \system-U} for \system-Uniform, {\bf MADAR$^\theta$-R} for MADAR$^\theta$-Ratio, and {\bf MADAR$^\theta$-U} for MADAR$^\theta$-Uniform.

For all three scenarios, we compare \system\ against widely studied replay-based continual learning (CL) techniques, including experience replay (ER)\cite{er}, average gradient episodic memory (AGEM)\cite{agem}, deep generative replay (GR)\cite{gr}, Replay-through-Feedback (RtF)\cite{rtf}, and Brain-inspired Replay (BI-R)\cite{BIR}. Additionally, we evaluate \system\ against iCaRL\cite{icarl}, a replay-based method specifically designed for Class-IL. For the Class-IL and Task-IL scenarios, we additionally compare \system\ with Task-specific Attention Modules in Lifelong Learning (TAMiL)\cite{tamil}. Furthermore, we benchmark MADAR against MalCL\cite{malcl}, a method specifically designed for Class-IL. Notably, most recent work focuses primarily on Class-IL and Task-IL scenarios, limiting direct comparisons in the Domain-IL scenario. In our results tables, the best-performing methods and those within the error margin of the top results are highlighted.

\subsection{Domain-IL}
\label{domainilexps}


In EMBER, we have 12 tasks, each representing the monthly data distribution spanning January--December 2018. Our results, detailed in Table~\ref{tab:ember_DIL}, provide a comprehensive view of each method's performance, reported as the average accuracy over all tasks $\mathbf{\overline{AP}}$. 

The informal lower and upper performance bounds for this configuration are approximated by the \textit{None} and \textit{Joint} methods, achieving $\mathbf{\overline{AP}}$ scores of 93.1\% and 96.4\%, respectively. Meanwhile, \textit{GRS} serves as a strong baseline, providing unbiased sampling without incorporating sample heterogeneity awareness.


At a lower budget of 1K, \system-R, \system-U, and MADAR$^\theta$-R exhibit competitive performance, all achieving $\mathbf{\overline{AP}}$ of over $93.6$\%, significantly outperforming prior approaches. This highlights their ability to effectively utilize limited resources. In particular, \system-R achieves the highest accuracy at this budget, with $\mathbf{\overline{AP}}$ of $93.7\%$. As the memory budget increases, the performance of all \system\ and MADAR$^\theta$ variants improves steadily. At a budget of 200K, \system-R and MADAR$^\theta$-R achieve an impressive $\mathbf{\overline{AP}}$ of $96.0\%$ and $96.1\%$, respectively, closely approaching the $96.4\%$ achieved by the \textit{Joint} baseline, which utilizes over 670K samples. Uniform strategies, including \system-U and MADAR$^\theta$-U, are only slightly behind, with $\mathbf{\overline{AP}}$ values of $95.5\%$ and $95.6\%$, respectively.


\begin{table*}[!t]
\small
\centering
\caption{Summary of Results for EMBER Class-IL Experiments.}
\vspace{-0.3cm}
\label{tab:ember_CIL}
\begin{tabular}{p{1.1cm}|l|c|c|c|c|c|c|c} 


\multirow{2}{*}{\textbf{Group}} & \multirow{2}{*}{\textbf{Method}} & \multicolumn{7}{c}{\textbf{Budget}} \\ \cline{3-9}

&  & 100 & 500 & 1K & 5K & 10K & 15K & 20K \\ \midrule

\multirow{3}{*}{Baselines} 
& Joint  & \multicolumn{7}{c}{86.5$\pm$0.4} \\ 
& None   & \multicolumn{7}{c}{26.5$\pm$0.2} \\ 
& GRS    & 51.9$\pm$0.4 & 70.3$\pm$0.5 & 75.4$\pm$0.7 & 82.0$\pm$0.2 & 83.5$\pm$0.1 & 84.3$\pm$0.3 & 84.6$\pm$0.2 \\ \midrule

\multirow{6}{*}{\parbox{0.7cm}{Prior \\ Work}} 
& TAMiL~\cite{tamil}  & 32.2$\pm$0.3 & 33.1$\pm$0.2 & 35.3$\pm$0.2 & 36.7$\pm$0.1 & 38.2$\pm$0.3 & 37.2$\pm$0.2 & 38.8$\pm$0.2 \\ 
& iCaRL~\cite{icarl}  & 53.9$\pm$0.7 & 58.7$\pm$0.7 & 60.0$\pm$1.0 & 63.9$\pm$1.2 & 64.6$\pm$0.8 & 65.5$\pm$1.0 & 66.8$\pm$1.1 \\ 
& ER~\cite{er}     & 27.5$\pm$0.1 & 27.8$\pm$0.1 & 28.0$\pm$0.1 & 27.9$\pm$0.1 & 28.0$\pm$0.1 & 28.0$\pm$0.1 & 28.2$\pm$0.1 \\ 
& AGEM~\cite{agem}   & 27.3$\pm$0.1 & 27.4$\pm$0.1 & 27.7$\pm$0.1 & 28.5$\pm$0.1 & 28.2$\pm$0.1 & 28.3$\pm$0.1 & 28.2$\pm$0.1 \\ 
& GR~\cite{gr}     & \multicolumn{7}{c}{26.8$\pm$0.2} \\ 
& RtF~\cite{rtf}   & \multicolumn{7}{c}{26.5$\pm$0.1} \\ 
& BI-R~\cite{BIR}   & \multicolumn{7}{c}{26.9$\pm$0.1} \\ 
& MalCL~\cite{malcl}   & \multicolumn{7}{c}{54.5$\pm$0.3} \\ 
\midrule

\multirow{4}{*}{\system} 
& \system-R & \textbf{68.0$\pm$0.4} & 73.6$\pm$0.2 & 76.0$\pm$0.3 & 81.5$\pm$0.2 & 83.2$\pm$0.2 & 83.8$\pm$0.2 & 84.0$\pm$0.2 \\ 
& \system-U & 66.4$\pm$0.4 & \textbf{76.5$\pm$0.2} & \textbf{79.4$\pm$0.4} & \textbf{83.8$\pm$0.2} & \textbf{84.8$\pm$0.1} & \textbf{85.5$\pm$0.1} & \textbf{85.8$\pm$0.3} \\ \cline{2-9}
& MADAR$^{\theta}$-R & {\bf 67.9$\pm$0.3} & 72.7$\pm$0.5 & 72.7$\pm$0.5 & 81.7$\pm$0.2 & 83.2$\pm$0.1 & 83.9$\pm$0.1 & 84.5$\pm$0.2 \\ 
& MADAR$^{\theta}$-U & 67.5$\pm$0.3 & {\bf 76.4$\pm$0.4} & {\bf 78.5$\pm$0.4} & {\bf 84.1$\pm$0.1} & {\bf 85.3$\pm$0.1} & {\bf 85.8$\pm$0.2} & {\bf 86.2$\pm$0.2} \\ 

\bottomrule

\end{tabular}
\vspace{-0.2cm}
\end{table*}

\begin{figure}[!t]
    \centering
    \begin{subfigure}{0.485\linewidth}
        \centering
        \includegraphics[width=1.0\linewidth]{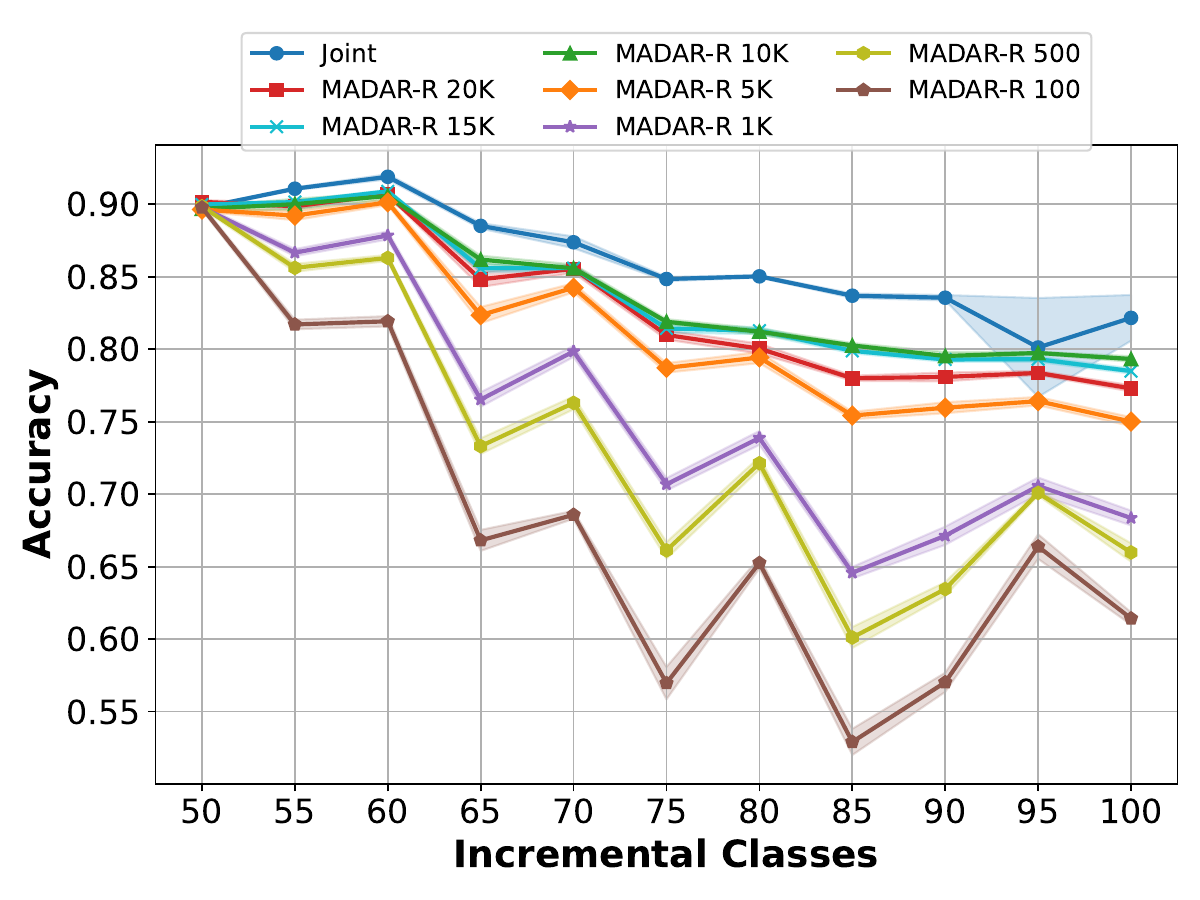}
        \label{fig:EMBER_CIL_IFS_R}
        \vspace{-0.4cm}
        \caption{MADAR Ratio}
    \end{subfigure}
    \hfill
    \begin{subfigure}{0.485\linewidth}
        \centering
        \includegraphics[width=1.0\linewidth]{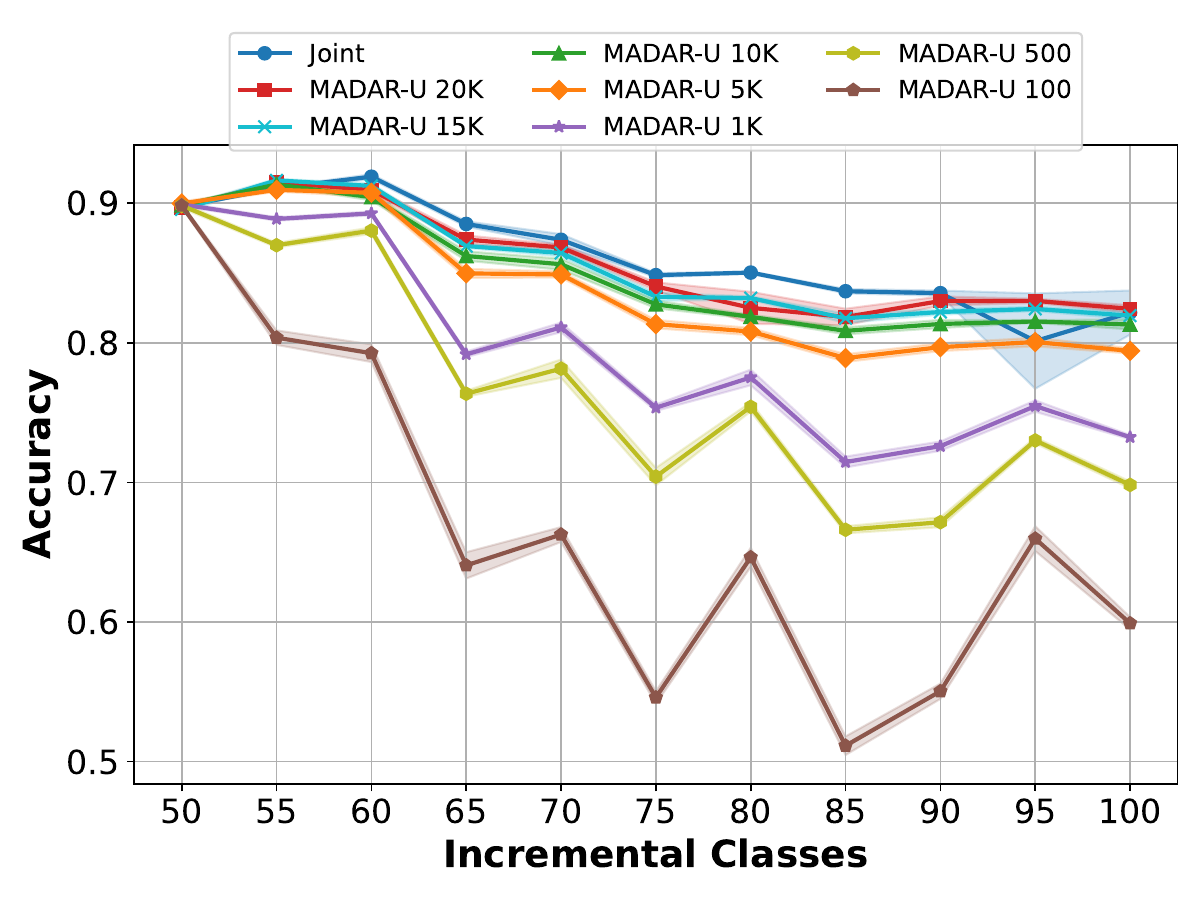}
        \label{fig:EMBER_CIL_IFS_U}
        \vspace{-0.4cm}
        \caption{MADAR Uniform}
    \end{subfigure}
    \vfill
    \begin{subfigure}{0.485\linewidth}
        \centering
        \includegraphics[width=1.0\linewidth]{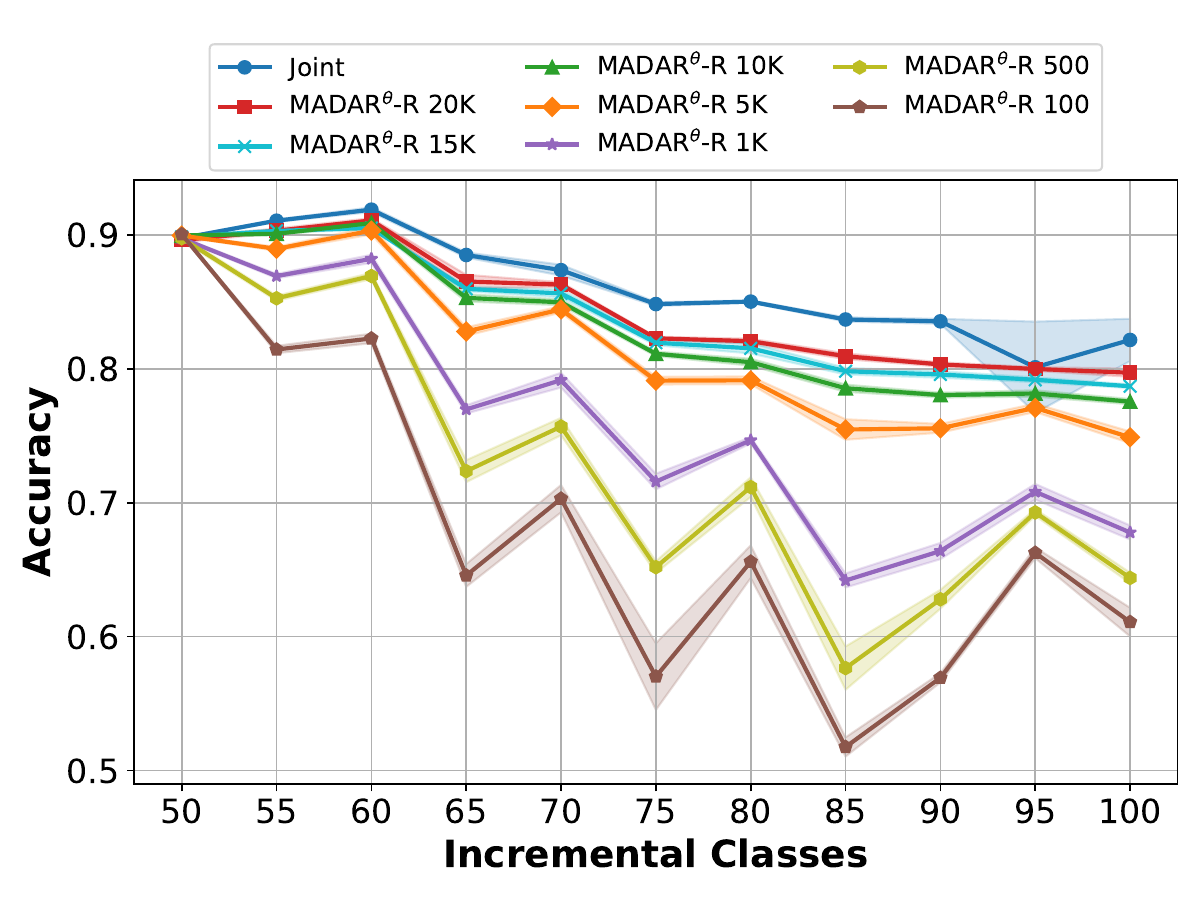}
        \label{fig:EMBER_CIL_AWS_R}
        \vspace{-0.4cm}
        \caption{MADAR$^\theta$ Ratio}
    \end{subfigure}
    \hfill
    \begin{subfigure}{0.485\linewidth}
        \centering
        \includegraphics[width=1.0\linewidth]{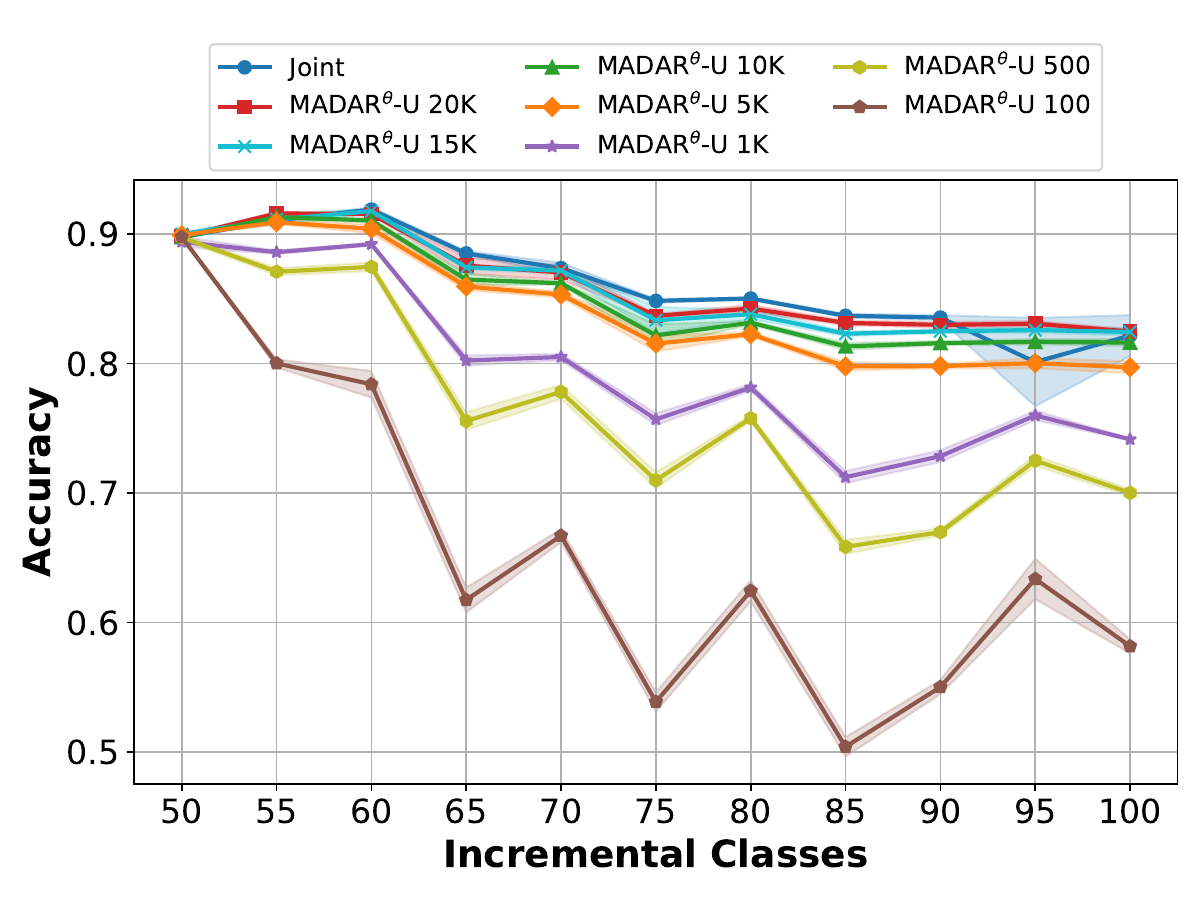}
        \label{fig:EMBER_CIL_AWS_U}
        \vspace{-0.4cm}
        \caption{MADAR$^\theta$ Uniform}
    \end{subfigure}

    \caption{EMBER Class-IL: Comparison of the MADAR-R, MADAR-U, MADAR$^\theta$-R, and MADAR$^\theta$-U with Joint baseline.}
    \label{fig:ember_CIL}
    \vspace{-0.3cm}
\end{figure}

For the experiments with AZ-Domain, we consider 9 tasks, each representing a yearly data distribution from 2008 to 2016. The performance of each method is presented in Table~\ref{tab:az_DIL} as $\mathbf{\overline{AP}}$ and compared to two baselines: \textit{None}, which achieves $94.4\%$, and \textit{Joint}, which reaches $97.3\%$. 

Similar to the results observed with EMBER, our MADAR techniques consistently outperform prior methods such as ER, AGEM, GR, RtF, and BI-R across all budget levels. For lower budgets, such as 1K, \system-R achieves $\mathbf{\overline{AP}}$ of $95.8\%$ and coming within 1.5\% of the \textit{Joint} baseline.

At higher budgets, ranging from 100K to 400K, \system-R continues to demonstrate high $\mathbf{\overline{AP}}$ scores of up to $97.0\%$, closely matching GRS and only marginally below the \textit{Joint} baseline, which requires significantly more training samples (680K). Notably, MADAR$^\theta$-R exhibits comparable performance, reaching a peak $\mathbf{\overline{AP}}$ of $97.2\%$ at the highest budget level, further underscoring the efficacy of our distribution-aware approach.




In summary, these results empirically demonstrate the effectiveness of MADAR's distribution-aware sample selection in enhancing the efficiency and accuracy of malware classification in Domain-IL scenarios. \system-R and MADAR$^\theta$-R, in particular, consistently either yield on-par or outperform GRS while delivering significant improvements over prior methods.

\begin{table*}[!t]
\small
\centering
\caption{Summary of Results for AZ Class-IL Experiments.}
\vspace{-0.3cm}
\label{tab:az_CIL}
\begin{tabular}{p{1.1cm}|l|c|c|c|c|c|c|c} 


\multirow{2}{*}{\textbf{Group}} & \multirow{2}{*}{\textbf{Method}} & \multicolumn{7}{c}{\textbf{Budget}} \\ \cline{3-9}

&  & 100 & 500 & 1K & 5K & 10K & 15K & 20K \\ \midrule

\multirow{3}{*}{Baselines} 
& Joint  & \multicolumn{7}{c}{94.2$\pm$0.1} \\ 
& None   & \multicolumn{7}{c}{26.4$\pm$0.2} \\ 
& GRS    & 43.8$\pm$0.7 & 62.9$\pm$0.8 & 70.2$\pm$0.4 & 83.0$\pm$0.3 & 86.4$\pm$0.2 & 88.2$\pm$0.2 & 89.1$\pm$0.2 \\ \midrule

\multirow{6}{*}{\parbox{0.7cm}{Prior \\ Work}} 
& TAMiL~\cite{tamil}  & 53.4$\pm$0.3 & 55.2$\pm$0.3 & 57.6$\pm$0.3 & 60.8$\pm$0.2 & 63.5$\pm$0.1 & 65.3$\pm$0.5 & 67.7$\pm$0.3 \\ 
& iCaRL~\cite{icarl}  & 43.6$\pm$1.2 & 54.9$\pm$1.0 & 61.7$\pm$0.7 & 77.2$\pm$0.4 & 81.5$\pm$0.6 & 83.4$\pm$0.5 & 84.6$\pm$0.5 \\ 
& ER~\cite{er}     & 50.8$\pm$0.7 & 58.3$\pm$0.6 & 58.9$\pm$0.2 & 59.2$\pm$0.8 & 62.9$\pm$0.7 & 63.1$\pm$0.5 & 64.2$\pm$0.4 \\ 
& AGEM~\cite{agem}   & 27.3$\pm$0.7 & 28.0$\pm$1.4 & 27.1$\pm$0.3 & 28.0$\pm$0.6 & 28.2$\pm$1.0 & 29.8$\pm$2.6 & 28.0$\pm$0.8 \\ 
& GR~\cite{gr}     & \multicolumn{7}{c}{22.7$\pm$0.3} \\ 
& RtF~\cite{rtf}    & \multicolumn{7}{c}{22.9$\pm$0.3} \\ 
& BI-R~\cite{BIR}   & \multicolumn{7}{c}{23.4$\pm$0.2} \\ 
& MalCL~\cite{malcl}   & \multicolumn{7}{c}{59.8$\pm$0.4} \\ 
\midrule

\multirow{4}{*}{\system} 
& \system-R & \textbf{59.4$\pm$0.6} & 67.8$\pm$0.9 & 71.9$\pm$0.5 & 82.9$\pm$0.2 & 86.3$\pm$0.1 & 88.2$\pm$0.2 & 89.1$\pm$0.1 \\ 
& \system-U & 57.3$\pm$0.5 & \textbf{70.4$\pm$0.4} & \textbf{76.2$\pm$0.2} & \textbf{86.8$\pm$0.1} & \textbf{89.8$\pm$0.1} & \textbf{91.0$\pm$0.1} & \textbf{91.5$\pm$0.1} \\ \cline{2-9}
& MADAR$^{\theta}$-R & {\bf 58.8$\pm$0.3} & 66.2$\pm$0.7 & 71.0$\pm$0.7 & 81.2$\pm$0.3 & 85.1$\pm$0.2 & 86.9$\pm$0.2 & 88.1$\pm$0.1 \\ 
& MADAR$^{\theta}$-U & 58.5$\pm$0.7 & {\bf 70.1$\pm$0.2} & {\bf 74.7$\pm$0.2} & {\bf 85.5$\pm$0.1} & {\bf 88.7$\pm$0.1} & {\bf 90.3$\pm$0.2} & {\bf 90.7$\pm$0.1} \\ 

\bottomrule

\end{tabular}
\vspace{-0.2cm}
\end{table*}

\begin{figure}[!t]
    \centering
    \begin{subfigure}{0.485\linewidth}
        \centering
        \includegraphics[width=1.0\linewidth]{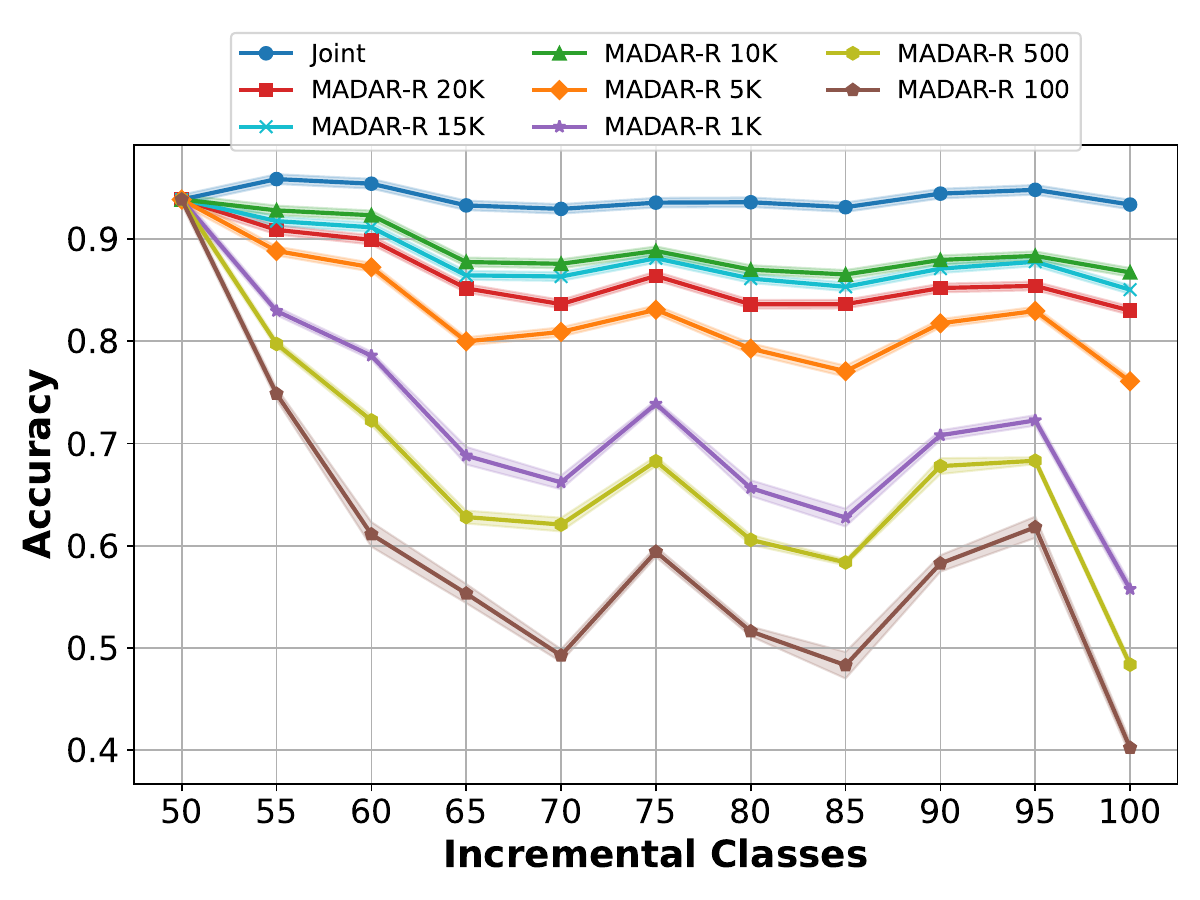}
        \label{fig:AZ_CIL_IFS_R}
        \vspace{-0.4cm}
        \caption{MADAR Ratio}
    \end{subfigure}
    \hfill
    \begin{subfigure}{0.485\linewidth}
        \centering
        \includegraphics[width=1.0\linewidth]{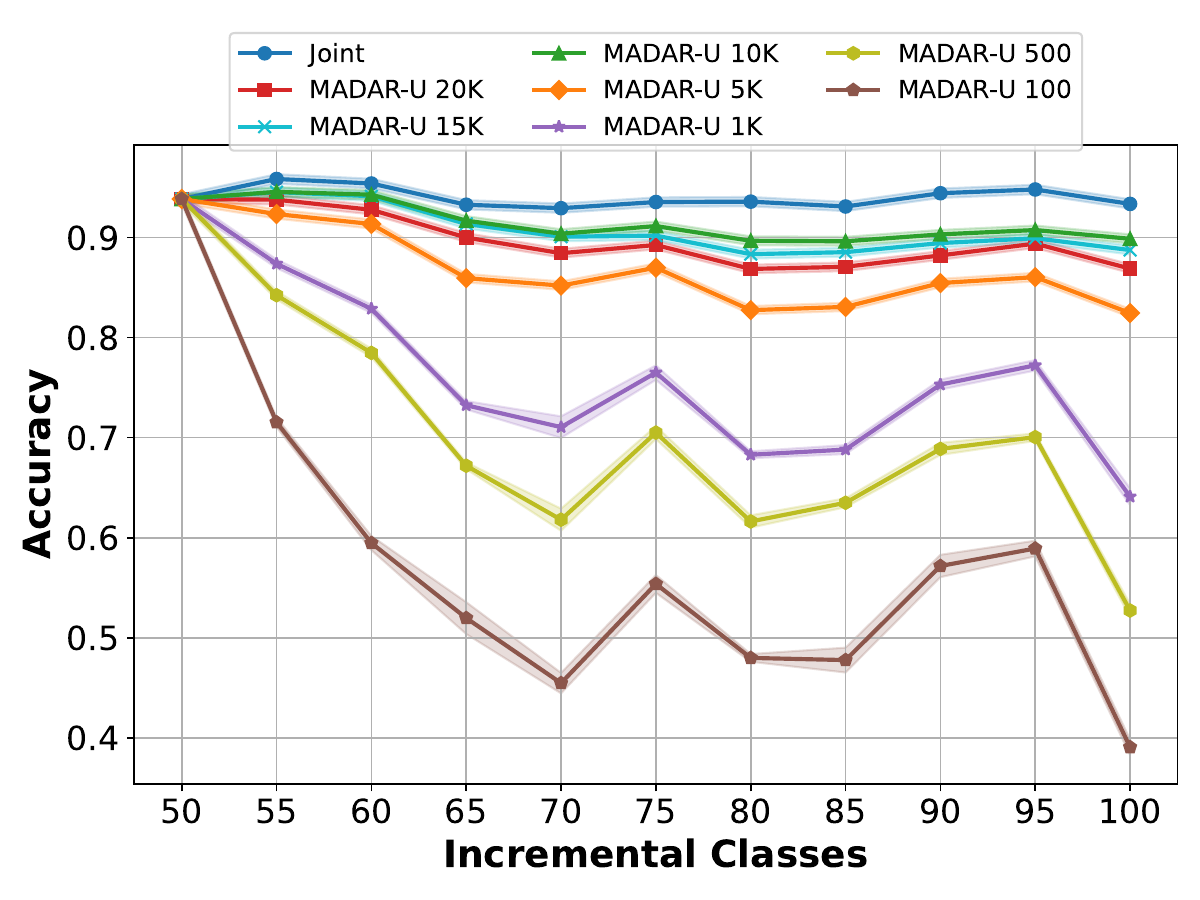}
        \label{fig:AZ_CIL_IFS_U}
        \vspace{-0.4cm}
        \caption{MADAR Uniform}
    \end{subfigure}
    \vfill
    \begin{subfigure}{0.485\linewidth}
        \centering
        \includegraphics[width=1.0\linewidth]{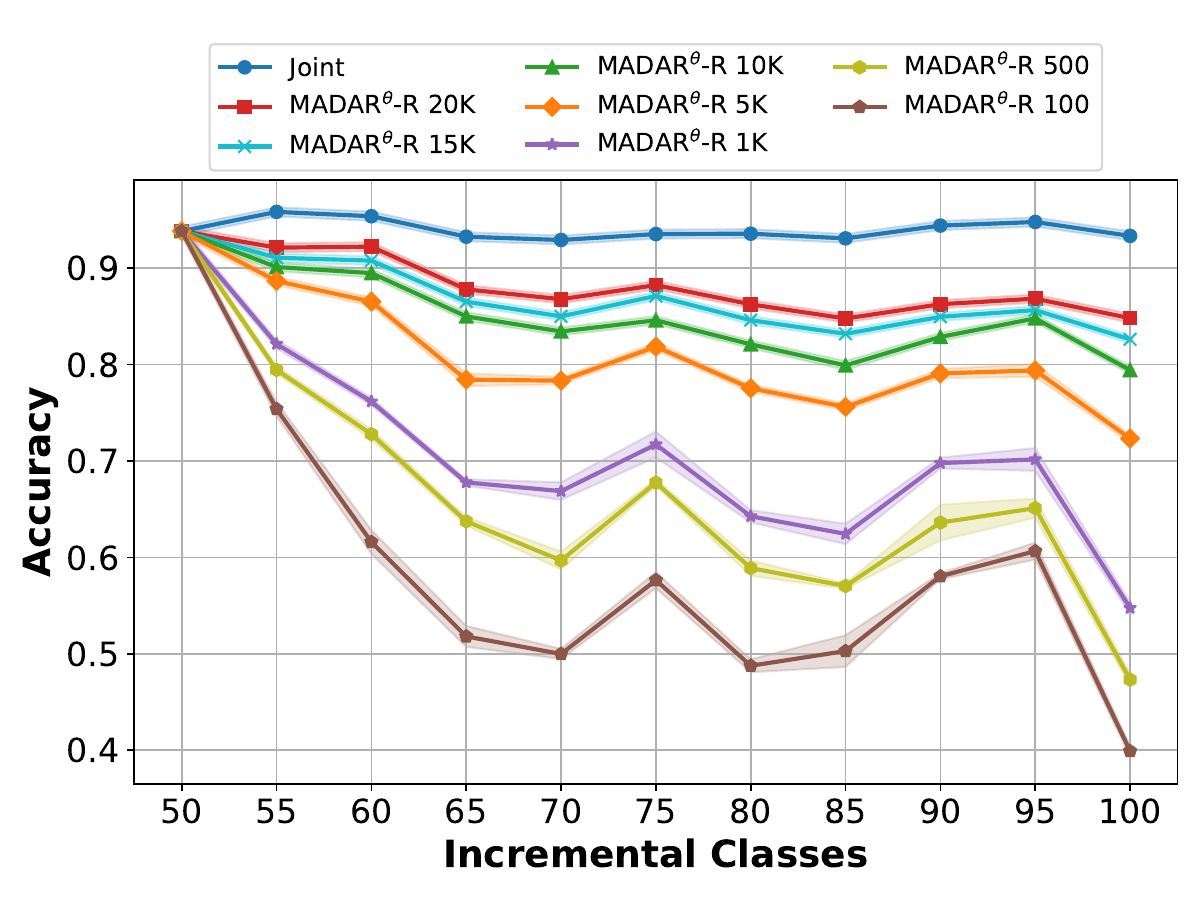}
        \label{fig:AZ_CIL_AWS_R}
        \vspace{-0.4cm}
        \caption{MADAR$^\theta$ Ratio}
    \end{subfigure}
    \hfill
    \begin{subfigure}{0.485\linewidth}
        \centering
        \includegraphics[width=1.0\linewidth]{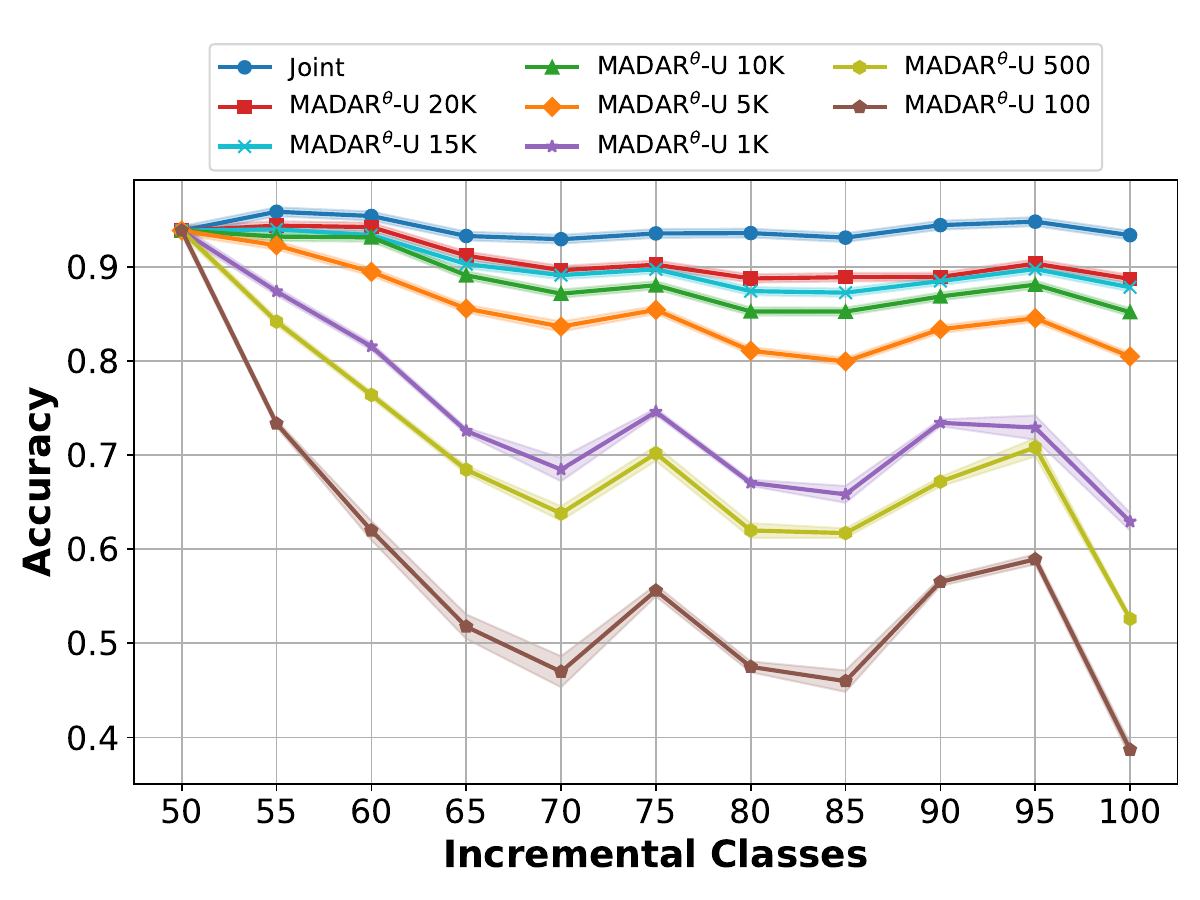}
        \label{fig:AZ_CIL_AWS_U}
        \vspace{-0.4cm}
        \caption{MADAR$^\theta$ Uniform}
    \end{subfigure}

    \caption{AZ Class-IL: Comparison of the MADAR-R, MADAR-U, MADAR$^\theta$-R, and MADAR$^\theta$-U with Joint baseline.}
    \label{fig:az_CIL}
    \vspace{-0.3cm}
\end{figure}

\subsection{Class-IL}
\label{classilexps}

In this set of experiments with EMBER, we consider 11 tasks, starting with 50 classes (representing distinct malware families) in the initial task, and incrementally adding five new classes in each subsequent task. Table~\ref{tab:ember_CIL} presents the performance of each method, measured by average accuracy $\mathbf{\overline{AP}}$. The \textit{None} and \textit{Joint} baselines achieve $\mathbf{\overline{AP}}$ values of $26.5\%$ and $86.5\%$, respectively, providing informal lower and upper bounds. Figure~\ref{fig:ember_CIL} illustrates the progression of average accuracy across tasks, showing how the \system\ and MADAR$^\theta$ methods compare to the \textit{Joint} baseline.

At a very low budget of just 100 samples, \system-R achieves a notable $\mathbf{\overline{AP}}$ of $68.0\%$, outperforming GRS and prior methods by a significant margin. As the budget increases, \system-U emerges as the top performer, achieving $\mathbf{\overline{AP}}$ values of $76.5\%$ and $79.4\%$ at 1K and 10K budgets, respectively, surpassing all other methods, including GRS. 


At higher budgets, \system-U and MADAR$^\theta$-U continue to excel, with MADAR$^\theta$-U achieving the best results overall. At a 20K budget, MADAR$^\theta$-U reaches an $\mathbf{\overline{AP}}$ of $86.2\%$, nearly equaling the \textit{Joint} baseline, which uses over {\bf 150 times} more training samples. These results clearly demonstrate the effectiveness of MADAR's distribution-aware sample selection and the effectiveness of \system-U and MADAR$^\theta$-U in leveraging limited resources.

In contrast, prior methods such as ER, AGEM, GR, RtF, and BI-R fail to exceed 30\% $\mathbf{\overline{AP}}$, while more advanced techniques like TAMiL and MalCL achieve only $38.2\%$ and $54.8\%$, respectively. Even iCaRL, designed specifically for Class-IL, achieves only $64.6\%$, further highlighting the significant advantage of our approaches in the malware domain.



In the Class-IL setting of AZ-Class, we consider 11 tasks. The summary results of all experiments are provided in Table~\ref{tab:az_CIL}, with comparisons against the \textit{None} and \textit{Joint} baselines, which achieve $\mathbf{\overline{AP}}$ scores of $26.4\%$ and $94.2\%$, respectively. Figure~\ref{fig:az_CIL} illustrates the progression of average accuracy across tasks, showing how each method performs relative to the \textit{Joint} baseline.

As shown in Table~\ref{tab:az_CIL}, among the prior methods, iCaRL performs best across most budget configurations, outperforming techniques such as MalCL, TAMiL, ER, AGEM, GR, RtF, and BI-R. Therefore, we focus on comparing MADAR's performance with iCaRL. At a low budget of 100 samples, iCaRL and GRS achieve less than $44\%$ $\mathbf{\overline{AP}}$, while all MADAR methods surpass $57\%$. In particular, \system-R and MADAR$^\theta$-R achieve $\mathbf{\overline{AP}}$ scores of $59.4\%$ and $58.8\%$, respectively, highlighting their efficiency at low-resource levels.

As the budget increases, all methods improve, but \system-U consistently delivers the best results. At a budget of 1K, \system-U achieves the highest $\mathbf{\overline{AP}}$ at $70.4\%$, followed closely by MADAR$^\theta$-U at $70.1\%$. This trend continues as budgets increase, with \system-U outperforming all other methods, achieving $\mathbf{\overline{AP}}$ scores of $89.8\%$ at 10K and $91.5\%$ at 20K. Compared to GRS, which achieves $90.1\%$ at 20K, and iCaRL, which trails at $84.6\%$, \system-U demonstrates clear superiority. MADAR$^\theta$-U also performs GRS reaching $90.7\%$ at 20K.



In summary, our experiments demonstrate the effectiveness of \system's distribution-aware replay techniques in the Class-IL setting for both EMBER and AZ datasets. While GRS shows significant improvement with larger budgets, \system's uniform variants consistently outperform it across all budget levels. These results underscore \system's ability to mitigate catastrophic forgetting and enhance malware classification performance, even in resource-constrained environments.


\begin{table*}[!t]
\small
\centering
\caption{Summary of Results for EMBER Task-IL Experiments.}
\vspace{-0.3cm}
\label{tab:ember_TIL}
\begin{tabular}{p{1.1cm}|l|c|c|c|c|c|c|c} 


\multirow{2}{*}{\textbf{Group}} & \multirow{2}{*}{\textbf{Method}} & \multicolumn{7}{c}{\textbf{Budget}} \\ \cline{3-9}

&  & 100 & 500 & 1K & 5K & 10K & 15K & 20K \\ \midrule

\multirow{3}{*}{Baselines} 
& Joint  & \multicolumn{7}{c}{97.0$\pm$0.3} \\ 
& None   & \multicolumn{7}{c}{74.6$\pm$0.7} \\ 
& GRS    & 86.9$\pm$0.3 & 87.4$\pm$0.3 & 93.6$\pm$0.3 & 94.4$\pm$0.2 & 94.7$\pm$0.3 & 94.9$\pm$0.1 & 95.0$\pm$0.1 \\ \midrule

\multirow{6}{*}{\parbox{0.7cm}{Prior \\ Work}} 
& TAMiL~\cite{tamil}  & 72.8$\pm$0.1 & 81.5$\pm$0.3 & 86.9$\pm$0.2 & 88.1$\pm$0.3 & 90.3$\pm$0.1 & 93.2$\pm$0.3 & 94.2$\pm$0.7 \\ 
& ER~\cite{er}     & 67.4$\pm$0.3 & 84.9$\pm$0.2 & 89.5$\pm$0.5 & 93.9$\pm$0.2 & 94.8$\pm$0.2 & 95.2$\pm$0.1 & 95.4$\pm$0.1 \\ 
& AGEM~\cite{agem}   & 79.6$\pm$0.2 & 81.7$\pm$0.2 & 83.8$\pm$0.4 & 84.9$\pm$0.2 & 86.1$\pm$0.2 & 88.9$\pm$0.2 & 89.3$\pm$0.1 \\ 
& GR~\cite{gr}     & \multicolumn{7}{c}{79.8$\pm$0.3} \\ 
& RtF~\cite{rtf}    & \multicolumn{7}{c}{77.8$\pm$0.2} \\ 
& BI-R~\cite{BIR}   & \multicolumn{7}{c}{87.2$\pm$0.3} \\ \midrule

\multirow{4}{*}{\system} 
& \system-R & 92.1$\pm$0.2 & 92.3$\pm$0.9 & 93.8$\pm$0.2 & 94.2$\pm$0.1 & 94.8$\pm$0.2 & {\bf 95.7$\pm$0.2} & {\bf 95.6$\pm$0.1} \\ 
& \system-U & {\bf 93.4$\pm$0.2} & {\bf 93.7$\pm$0.3} & {\bf 93.9$\pm$0.3} & {\bf 94.8$\pm$0.2} & {\bf 95.6$\pm$0.1} & {\bf 95.7$\pm$0.1} & {\bf 95.8$\pm$0.2} \\ \cline{2-9}
& MADAR$^{\theta}$-R & {\bf 93.1$\pm$0.2} & {\bf 93.3$\pm$0.1} & {\bf 93.6$\pm$0.1} & 94.3$\pm$0.1 & 94.6$\pm$0.2 & 94.8$\pm$0.2 & 94.7$\pm$0.3 \\ 
& MADAR$^{\theta}$-U & {\bf 93.2$\pm$0.1} & 93.1$\pm$0.2 & {\bf 93.8$\pm$0.2} & {\bf 94.4$\pm$0.1} & {\bf 94.8$\pm$0.1} & {\bf 95.3$\pm$0.2} & {\bf 95.5$\pm$0.3} \\ 

\bottomrule

\end{tabular}
\vspace{-0.3cm}
\end{table*}

\begin{figure}[!t]
    \centering
    \begin{subfigure}{0.49\linewidth}
        \centering
        \includegraphics[width=1.0\linewidth]{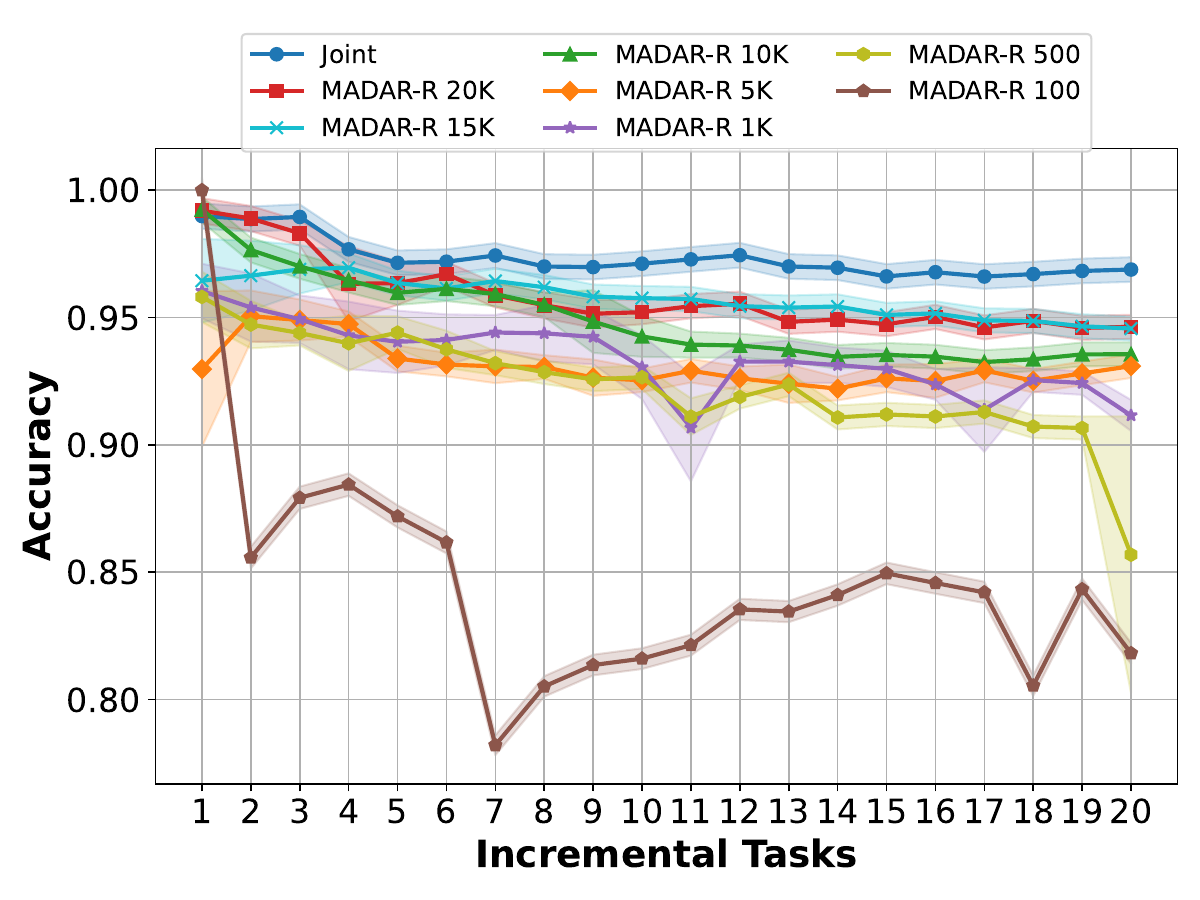}
        \label{fig:EMBER_TIL_IFS_R}
        \vspace{-0.4cm}
        \caption{MADAR Ratio}
    \end{subfigure}
    \hfill
    \begin{subfigure}{0.49\linewidth}
        \centering
        \includegraphics[width=1.0\linewidth]{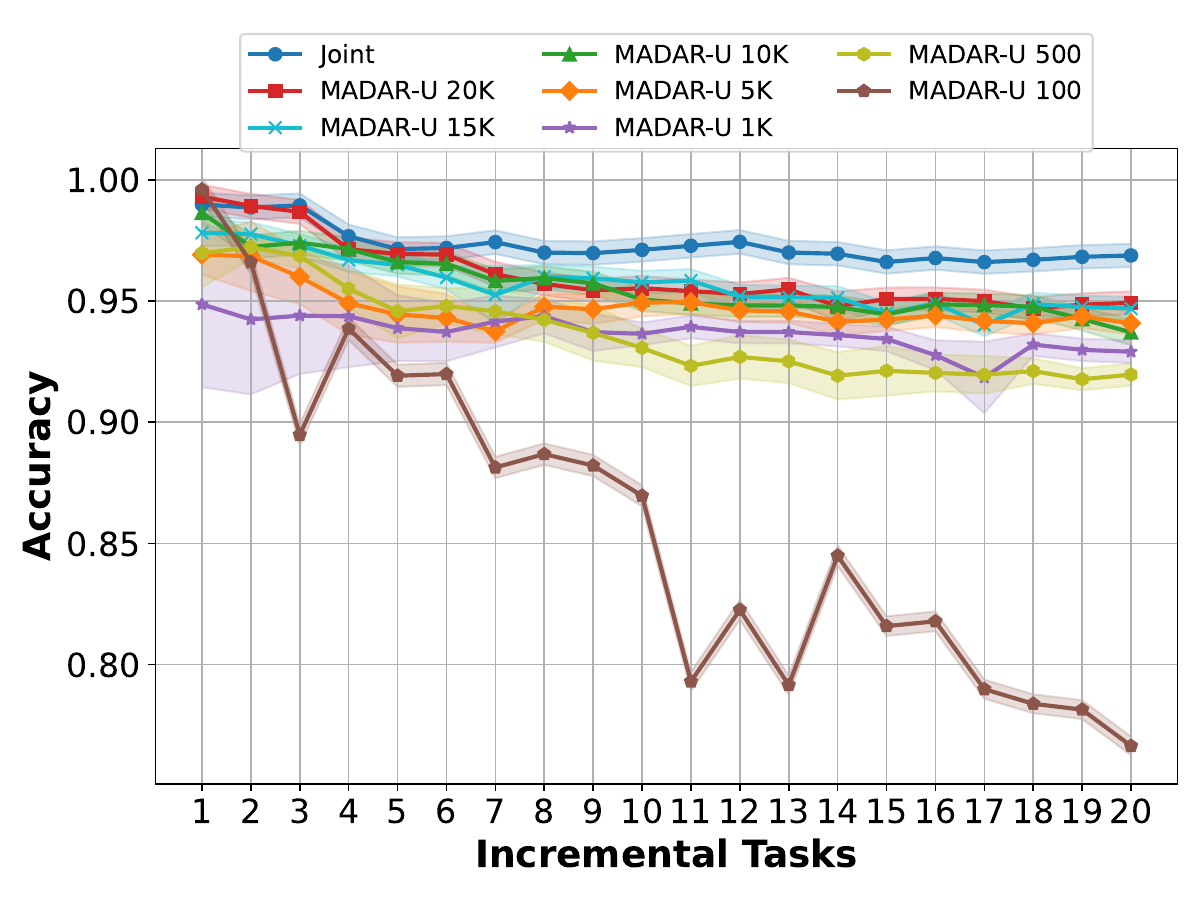}
        \label{fig:EMBER_TIL_IFS_U}
        \vspace{-0.4cm}
        \caption{MADAR Uniform}
    \end{subfigure}
    \vfill
    \begin{subfigure}{0.485\linewidth}
        \centering
        \includegraphics[width=1.0\linewidth]{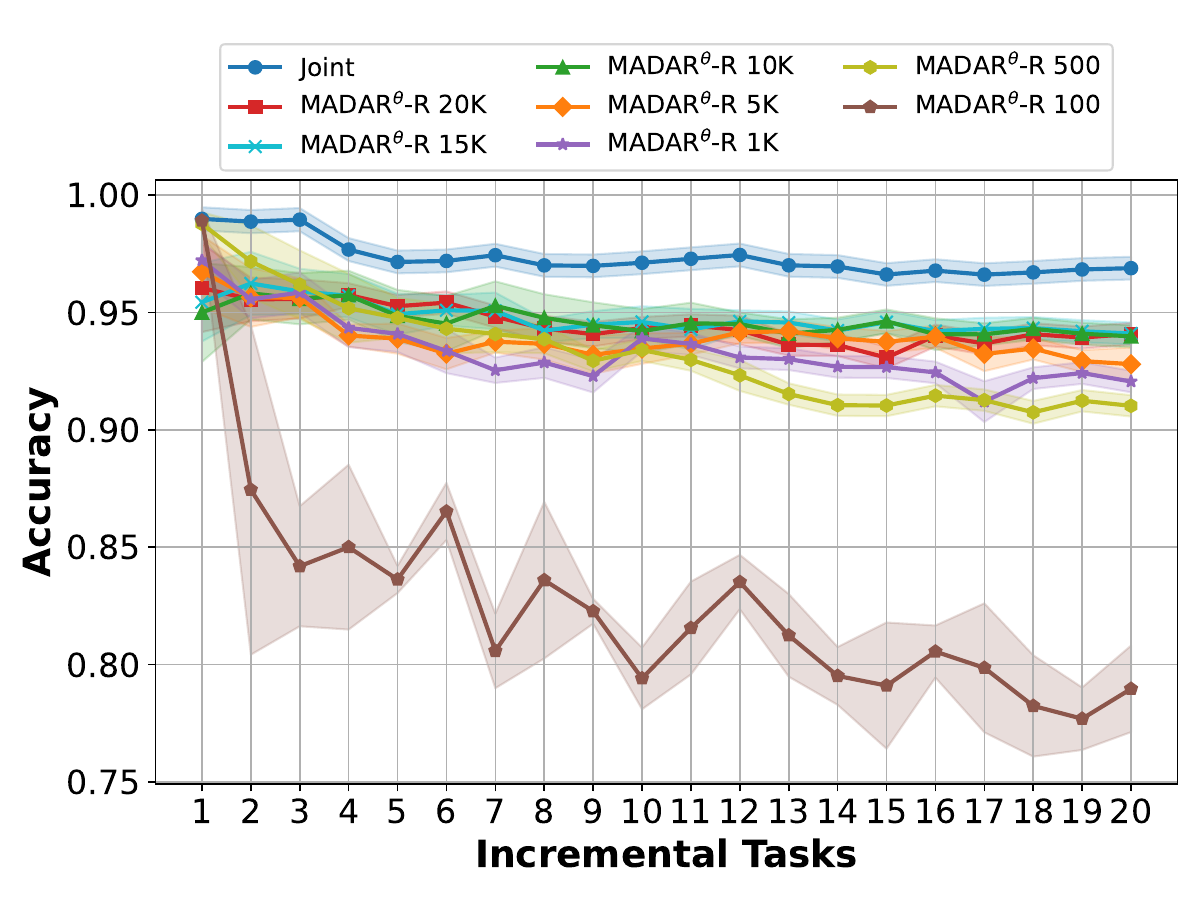}
        \label{fig:EMBER_TIL_AWS_R}
        \vspace{-0.4cm}
        \caption{MADAR$^\theta$ Ratio}
    \end{subfigure}
    \hfill
    \begin{subfigure}{0.485\linewidth}
        \centering
        \includegraphics[width=1.0\linewidth]{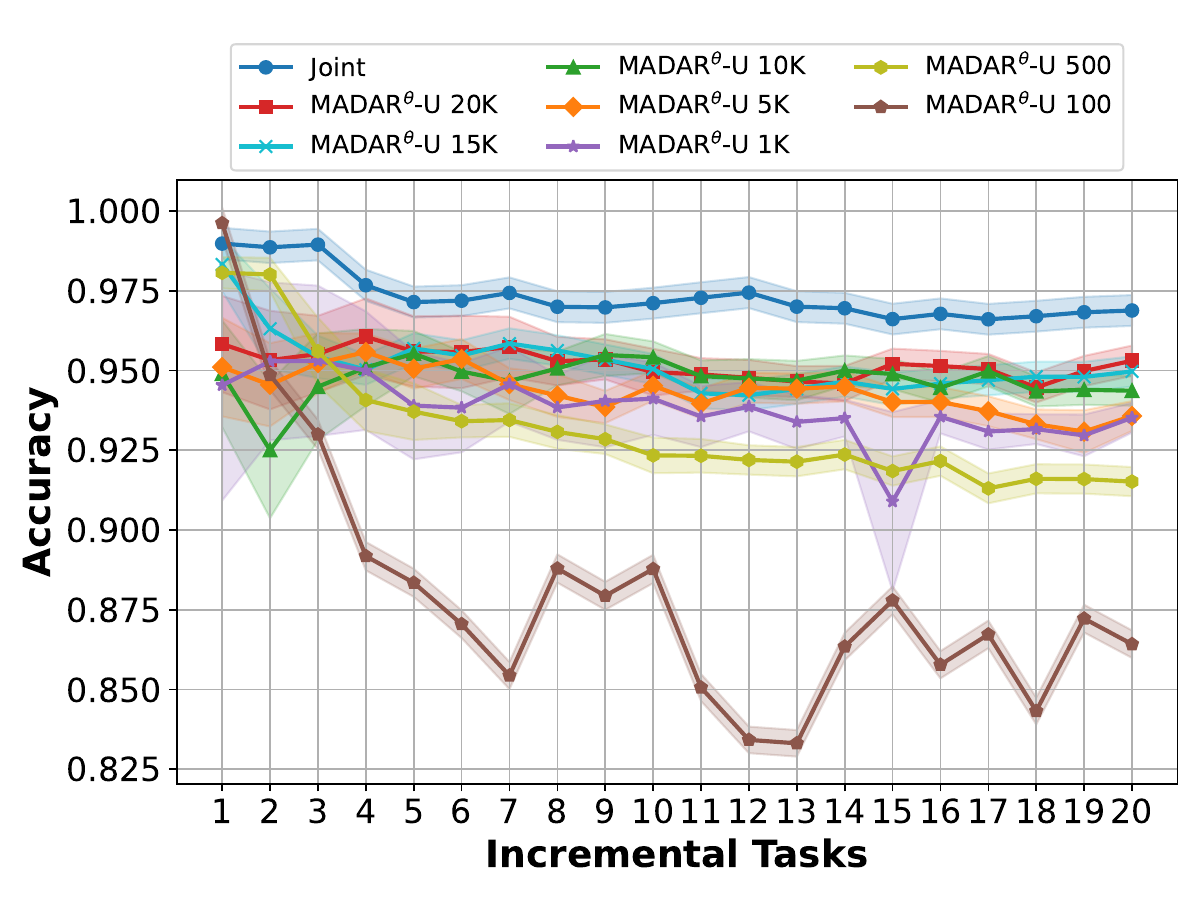}
        \label{fig:EMBER_TIL_AWS_U}
        \vspace{-0.4cm}
        \caption{MADAR$^\theta$ Uniform}
    \end{subfigure}

    \caption{EMBER Task-IL: Comparison of the MADAR-R, MADAR-U, MADAR$^\theta$-R, and MADAR$^\theta$-U with Joint baseline.}
    \label{fig:ember_TIL}
    \vspace{-0.3cm}
\end{figure}

\subsection{Task-IL}
\label{taskilexps-ember}

In this set of experiments with EMBER, we consider 20 tasks, with 5 new classes added in each task. The summarized results are shown in Table~\ref{tab:ember_TIL}, where performance is reported as the average accuracy over all tasks ($\mathbf{\overline{AP}}$). It is worth noting that Task-IL is considered the easiest scenario in continual learning~\cite{van2022three, BIR}. The \textit{None} and \textit{Joint} methods serve as informal lower and upper bounds, achieving $\mathbf{\overline{AP}}$ scores of $74.6\%$ and $97\%$, respectively. Figure~\ref{fig:ember_TIL} visualizes the progression of average accuracy across tasks, highlighting comparisons with the \textit{Joint} baseline.

As shown in Table~\ref{tab:ember_TIL}, ER consistently outperforms TAMiL, A-GEM, GR, RtF, and BI-R across all budget configurations and even surpasses GRS in some cases. However, \system\ variants significantly outperform all prior methods, particularly under lower budget constraints (100–1K). For example, \system-U achieves the highest $\mathbf{\overline{AP}}$ of $93.4\%$ and $93.7\%$ at budgets of 100 and 1K, respectively, outperforming GRS and all other approaches. Similarly, MADAR$^\theta$-U performs competitively, with $\mathbf{\overline{AP}}$ of $93.2\%$ at a 100 budget and $93.8\%$ at 1K.

As the budget increases, the performance gap among \system, ER, and GRS narrows; however, \system\ variants continue to either outperform or match other techniques. Notably, the \system-U variant of MADAR achieves the best overall performance at a budget of 20K, attaining a $\mathbf{\overline{AP}}$ of $95.8\%$, which closely approaches the \textit{Joint} baseline. Similarly, \system-R yields $\mathbf{\overline{AP}}$ of $95.6\%$ at 20K.



\begin{figure}[!t]
    \centering
    \begin{subfigure}{0.49\linewidth}
        \centering
        \includegraphics[width=1.0\linewidth]{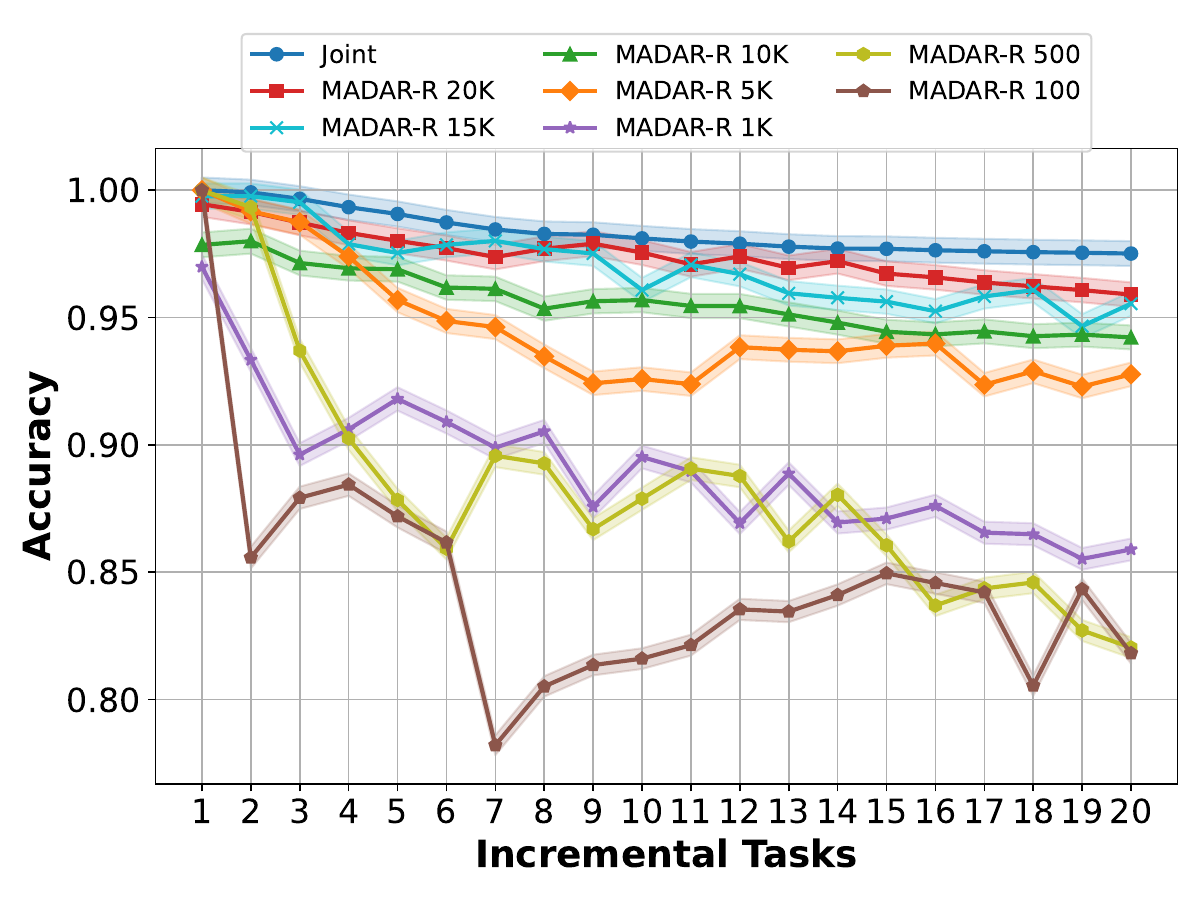}
        \label{fig:AZ_TIL_IFS_R}
        \vspace{-0.4cm}
        \caption{MADAR Ratio}
    \end{subfigure}
    \hfill
    \begin{subfigure}{0.49\linewidth}
        \centering
        \includegraphics[width=1.0\linewidth]{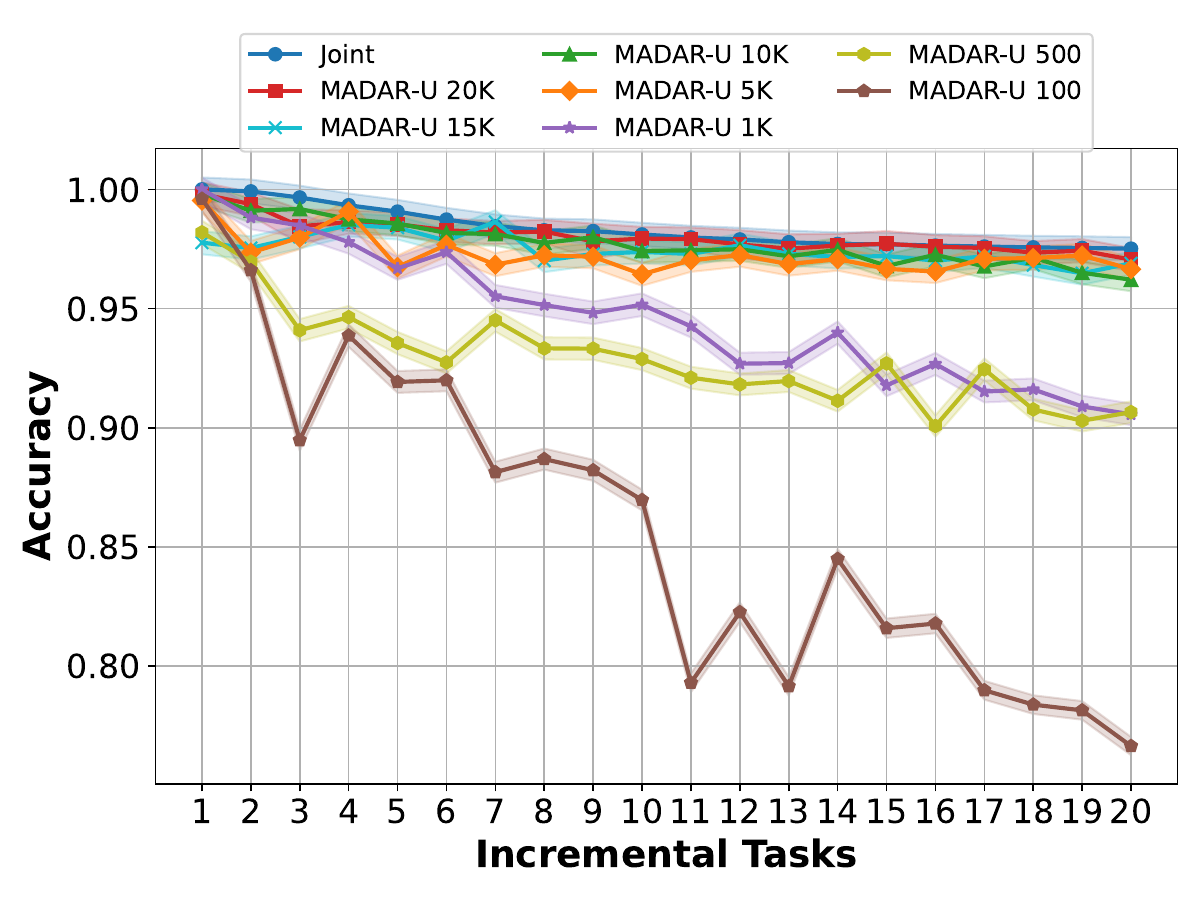}
        \label{fig:AZ_TIL_IFS_U}
        \vspace{-0.4cm}
        \caption{MADAR Uniform}
    \end{subfigure}
    \vfill
    \begin{subfigure}{0.45\linewidth}
        \centering
        \includegraphics[width=1.0\linewidth]{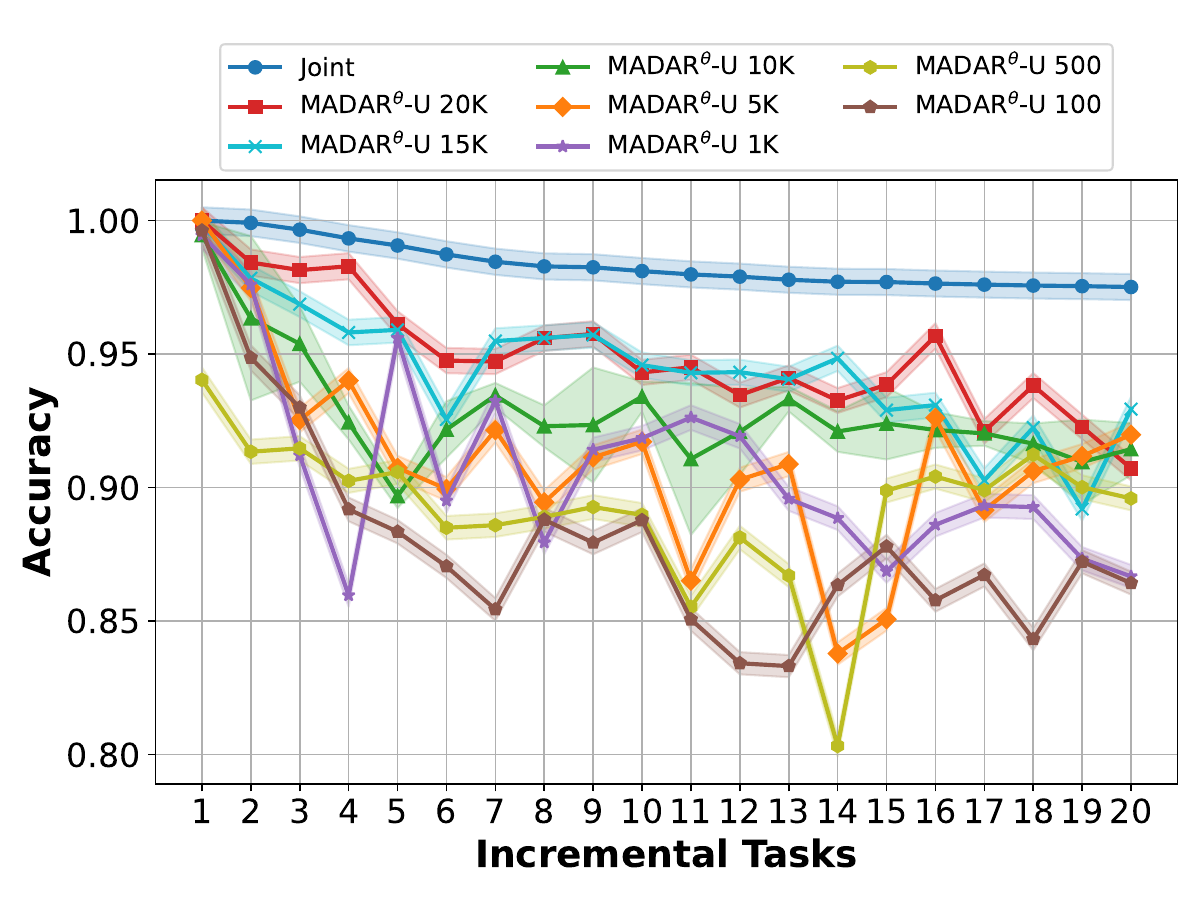}
        \label{fig:AZ_TIL_AWS_R}
        \vspace{-0.4cm}
        \caption{MADAR$^\theta$ Ratio}
    \end{subfigure}
    \hfill
    \begin{subfigure}{0.45\linewidth}
        \centering
        \includegraphics[width=1.0\linewidth]{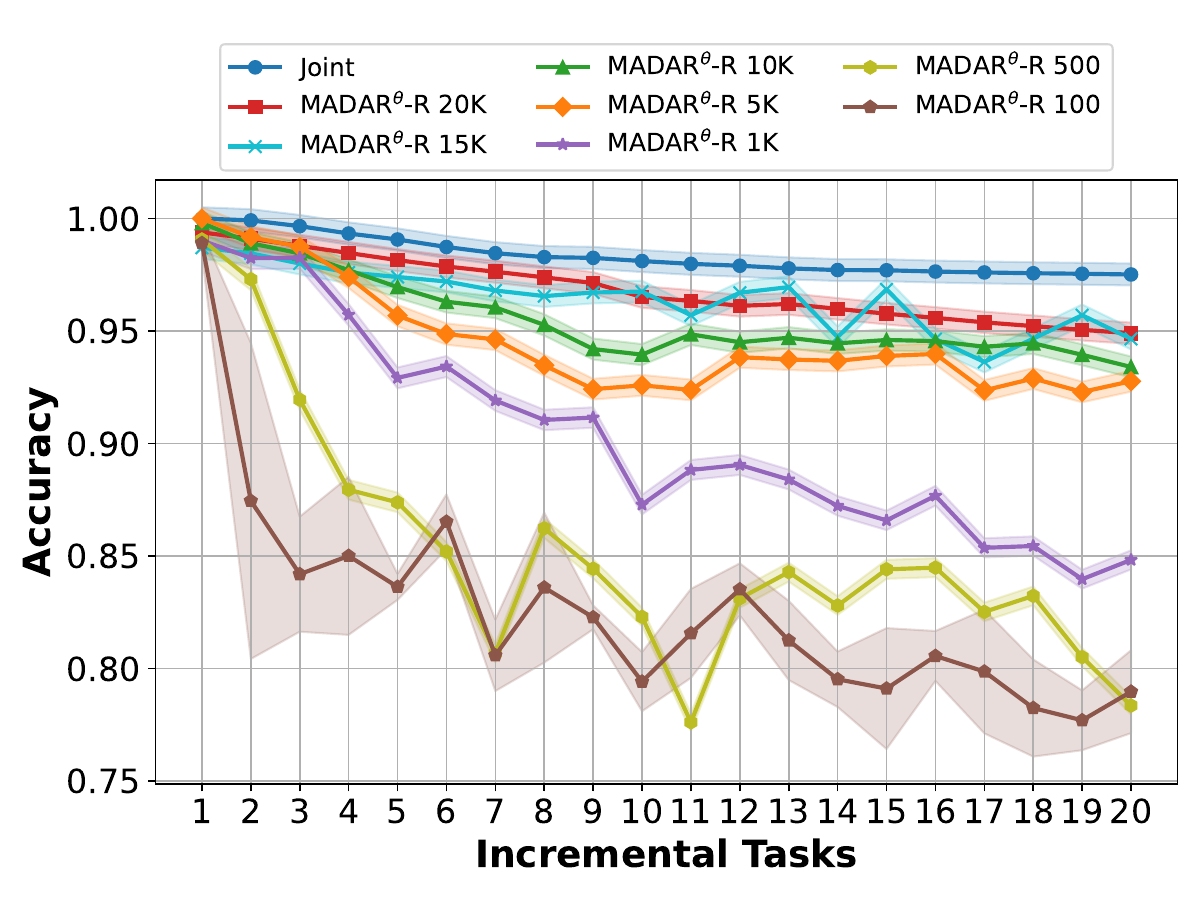}
        \label{fig:AZ_TIL_AWS_U}
        \vspace{-0.4cm}
        \caption{MADAR$^\theta$ Uniform}
    \end{subfigure}

    \caption{AZ Task-IL: Comparison of the MADAR-R, MADAR-U, MADAR$^\theta$-R, and MADAR$^\theta$-U with Joint baseline.}
    \label{fig:az_TIL}
    \vspace{-0.3cm}
\end{figure}

Task-IL for AZ consists of 20 tasks, each with 5 non-overlapping classes. The results are summarized in Table~\ref{tab:az_TIL} and benchmarked against the \textit{None} and \textit{Joint} baselines, which achieve $\mathbf{\overline{AP}}$ values of $74.5\%$ and $98.8\%$, respectively. Figure~\ref{fig:az_TIL} illustrates the progression of average accuracy across tasks, showing how each method performs relative to the \textit{Joint} baseline.

As seen in Table~\ref{tab:az_TIL}, ER consistently outperforms TAMiL, AGEM, GR, RtF, BI-R, and GRS across most budget configurations, making it a strong baseline for comparison. At a low budget of 100 samples, \system-U achieves $\mathbf{\overline{AP}}$ of $88.1\%$, which is 4.5\% higher than ER's performance. Similarly, MADAR$^\theta$-U demonstrates competitive performance, achieving $87.9\%$ at the same budget.

As the budget increases, \system-U continues to deliver the best performance, reaching $\mathbf{\overline{AP}}$ scores of $94.5\%$ at a 1K budget and $98.1\%$ at a 10K budget, outperforming all other methods, including ER and GRS. At the highest budget of 20K, \system-U achieves an $\mathbf{\overline{AP}}$ of $98.7\%$, surpassing ER by 1.2\% and nearly matching the \textit{Joint} baseline. Notably, MADAR$^\theta$-U also performs well, achieving $98.1\%$. In contrast, \system-R and MADAR$^\theta$-R perform slightly lower but remain competitive, with $\mathbf{\overline{AP}}$ values of $97.9\%$ and $96.9\%$ at a 20K budget, respectively. These results indicate that ratio-based methods, while effective, are slightly less robust than uniform sampling in this scenario.

In summary, \system-U and MADAR$^\theta$-U consistently demonstrate better performance across most of the budget levels, particularly excelling at low-resource settings and achieving near-optimal results at higher budgets. These findings underscore the effectiveness of \system\ framework in Task-IL scenarios and their ability to approach joint-level performance with significantly fewer resources.



\begin{table*}[!t]
\small
\centering
\caption{Summary of Results for AZ Task-IL Experiments.}
\vspace{-0.3cm}
\label{tab:az_TIL}
\begin{tabular}{p{1.1cm}|l|c|c|c|c|c|c|c} 


\multirow{2}{*}{\textbf{Group}} & \multirow{2}{*}{\textbf{Method}} & \multicolumn{7}{c}{\textbf{Budget}} \\ \cline{3-9}

&  & 100 & 500 & 1K & 5K & 10K & 15K & 20K \\ \midrule

\multirow{3}{*}{Baselines} 
& Joint  & \multicolumn{7}{c}{98.8$\pm$0.2} \\ 
& None   & \multicolumn{7}{c}{74.5$\pm$0.2} \\ 
& GRS    & 85.2$\pm$0.1 & 89.2$\pm$0.2 & 90.8$\pm$0.1 & 91.6$\pm$0.2 & 93.5$\pm$0.1 & 93.9$\pm$0.1 & 95.2$\pm$0.1 \\ \midrule

\multirow{6}{*}{\parbox{0.7cm}{Prior \\ Work}} 
& TAMiL  & 80.5$\pm$0.4 & 85.3$\pm$0.6 & 91.5$\pm$0.2 & 92.1$\pm$0.1 & 93.5$\pm$0.1 & 94.0$\pm$0.2 & 94.8$\pm$0.2 \\ 
& ER     & 83.6$\pm$0.2 & 90.2$\pm$0.1 & 92.3$\pm$0.3 & 95.6$\pm$0.1 & 96.2$\pm$0.1 & 96.8$\pm$0.2 & 97.5$\pm$0.2 \\ 
& AGEM   & 76.7$\pm$0.5 & 82.8$\pm$0.2 & 85.3$\pm$0.1 & 85.6$\pm$0.2 & 86.7$\pm$0.2 & 88.9$\pm$0.2 & 91.3$\pm$0.3 \\ 
& GR     & \multicolumn{7}{c}{75.6$\pm$0.2} \\ 
& RtF    & \multicolumn{7}{c}{74.2$\pm$0.3} \\ 
& BI-R   & \multicolumn{7}{c}{85.4$\pm$0.2} \\ \midrule

\multirow{4}{*}{\system} 
& \system-R & 86.0$\pm$0.3 & 90.3$\pm$0.2 & 92.4$\pm$0.1 & 95.8$\pm$0.2 & 96.7$\pm$0.1 & 97.1$\pm$0.1 & 97.9$\pm$0.2 \\ 
& \system-U & {\bf 88.1$\pm$0.3} & {\bf 92.9$\pm$0.2} & {\bf 94.5$\pm$0.3} & {\bf 97.2$\pm$0.2} & {\bf 98.1$\pm$0.1} & {\bf 98.5$\pm$0.1} & {\bf 98.7$\pm$0.1} \\ \cline{2-9}
& MADAR$^{\theta}$-R & 87.3$\pm$0.3 & {\bf 90.6$\pm$0.2} & 93.2$\pm$0.2 & 95.7$\pm$0.2 & 95.9$\pm$0.1 & 96.6$\pm$0.1 & 96.9$\pm$0.1 \\ 
& MADAR$^{\theta}$-U & {\bf 87.9$\pm$0.2} & {\bf 90.8$\pm$0.2} & {\bf 93.6$\pm$0.1} & {\bf 96.2$\pm$0.3} & {\bf 97.2$\pm$0.2} & {\bf 97.5$\pm$0.2} & {\bf 98.1$\pm$0.1} \\ 

\bottomrule

\end{tabular}
\vspace{-0.3cm}
\end{table*}

\subsection{Analysis and Discussion}\label{diss}

Our results demonstrate that MADAR yields markedly better performances compared to previous methods for both the EMBER and AZ datasets across all CL settings. This clearly indicates that distribution-aware replay is effective in preserving the stability of a CL-based system for malware classification, while prior CL techniques largely fail to achieve acceptable performance.

\paragraphX{\bf MADAR in low-budget settings.} In Domain-IL, MADAR achieves competitive performance even with a 1K budget, surpassing prior work by over 3 percentage points in EMBER and AZ. At higher budgets, ratio-based selection (\system-R and MADAR$^{\theta}$-R) achieves near Joint baseline performance (96.4\% in EMBER and 97.3\% in AZ) while using significantly fewer resources. This demonstrates MADAR’s efficiency in leveraging limited samples to achieve robust classification.

\paragraphX{\bf MADAR is both effective and scalable.} Traditional CL methods, including ER and AGEM, experience significant performance degradation as tasks increase. In contrast, MADAR maintains high accuracy across 20 Task-IL tasks, with \system-U achieving 95.8\% in EMBER and 98.7\% in AZ at a 20K budget, nearly matching the {\em Joint} baseline.

\paragraphX{\bf Ratio vs. Uniform Budgeting.} A consistent trend across our experiments is that ratio-based selection performs best in Domain-IL, whereas uniform-based selection is superior in Class-IL and Task-IL. MADAR$^{\theta}$-U reaches 91.5\% in AZ at 20K, significantly outperforming iCaRL and TAMiL. Furthermore, in EMBER, \system-U achieves near {\em Joint} baseline performance at just a 5K budget, underscoring the effectiveness of uniform selection in class-incremental settings. Intuitively, this makes sense because ratio budgeting for binary classification in the Domain-IL setting naturally captures the contributions of each family to the overall malware distribution. Additionally, since there are many small families in the Domain-IL datasets, uniformly sampling from them consumes budget while offering little improvement in malware coverage. In contrast, our Class-IL and Task-IL experiments perform classification across families, which is better supported by Uniform budgeting to maintain class balance and ensure coverage over all families. Moreover, in most settings we can leverage efficient representations using MADAR$^\theta$ to scale the approach regardless of feature dimension without significant loss of performance.

\paragraphX{\bf GRS remains a strong baseline at high budgets.} While MADAR consistently outperforms GRS in low-resource settings, GRS performs comparably at higher budgets, particularly in Domain-IL. This suggests that distribution-aware replay is most impactful when the number of available samples per class is limited, whereas uniform selection provides sufficient representation at larger budgets.

\section{Conclusion}

In this paper, we propose \system, a framework for distribution-aware replay in continual learning specially designed for the challenging setting of malware classification. Our comprehensive evaluation across Domain-IL, Class-IL, and Task-IL scenarios against Windows executable (EMBER) and Android application (AZ) datasets demonstrates that distribution-aware sampling is helpful for effective CL in malware classification. 
As malware and goodware continue to evolve, these insights steer continual learning towards strategic, resource-efficient methods, ensuring model effectiveness amid the constantly shifting landscape of cybersecurity threats.




\subsection*{\textbf{Acknowledgements}}
This research was funded in part by the National Science Foundation under Grant no. 2422241. The authors acknowledge Research Computing~\cite{rc-rit} at the Rochester Institute of Technology for providing computational resources and support.

\bibliographystyle{IEEEtran}
\bibliography{main}

\vfill

\end{document}